\PassOptionsToPackage{table}{xcolor}

\documentclass[letterpaper,twocolumn,10pt]{article}
\usepackage{usenix}

\usepackage{cite}
\usepackage{graphicx}
\usepackage{tikz}
\usepackage{hyperref}
\usepackage{etoolbox}
\usepackage{amsmath}
\usepackage{tabularx}
\usepackage{pifont}
\usepackage{booktabs}
\usepackage{tablefootnote}
\usepackage{subcaption}
\usepackage{listings}
\usepackage{multicol,multirow}
\usepackage{xcolor}

\usepackage{amssymb}
\usepackage{algorithm}
\usepackage{algpseudocode}
\usepackage{array}
\usepackage{enumitem}
\usepackage{tcolorbox}
\usepackage{paralist}
\usepackage{soul}

\newcommand{\cvegenie}{\textsc{CVE-Genie}}
\newcommand{\eager}{\textsc{EAGER}}

\newcommand{\totalcvesper}{51\%}
\newcommand{\totalcves}{428}
\newcommand{\totalcwes}{141}
\newcommand{\totalprojects}{267}
\newcommand{\totallangs}{22}
\newcommand{\avgcost}{\$2.77}

\newcommand{\cveprocessor}{\emph{Processor}}
\newcommand{\dataprocessor}{\emph{Data Processor}}
\newcommand{\knowledge}{\emph{Knowledge Builder}}
\newcommand{\project}{\emph{Builder}}
\newcommand{\predev}{\emph{Pre-Requisite Developer}}
\newcommand{\setupdev}{\emph{Setup Developer}}
\newcommand{\setupcritic}{\emph{Setup Critic}}
\newcommand{\exploit}{\emph{Exploiter}}
\newcommand{\exploitdev}{\emph{Exploit Developer}}
\newcommand{\exploitcritic}{\emph{Exploit Critic}}
\newcommand{\verifier}{\emph{CTF Verifier}}
\newcommand{\verifierdev}{\emph{Verifier Developer}}
\newcommand{\flagcheck}{\emph{Flag Checker}}
\newcommand{\verifiercritic}{\emph{Verifier Critic}}

\newcommand{\descr}[1]{\smallskip\noindent\textbf{#1}}
\newcommand{\descrit}[1]{\smallskip\noindent\emph{#1}}
\newcommand{\greentick}{\raisebox{-0.3em}{\textcolor{green!60!black}{\scalebox{1.5}{\ding{51}}}}}
\newcommand{\redcross}{\raisebox{-0.3em}{\textcolor{red}{\scalebox{1.5}{\ding{55}}}}}
\newcommand{\fullcircle}{\scalebox{1.35}{\ding{108}}}
\newcommand{\halfcircle}{\scalebox{1.3}{\ding{119}}}
\newcommand{\emptycircle}{\scalebox{1.3}{\ding{109}}}

\newcommand{\cirnum}[1]{%
    \begin{tikzpicture}[baseline=-0.7ex]
       \node[circle, draw, inner sep=1pt, thick] {\scriptsize \textbf{#1}};
    \end{tikzpicture}%
}
\robustify{\cirnum}

\definecolor{lightgraybox}{rgb}{0.95, 0.95, 0.95}
\definecolor{grayborder}{rgb}{0.3, 0.3, 0.3}

\tcbset{
  researchstyle/.style={
    colback=lightgraybox,
    colframe=grayborder,
    arc=3mm,
    boxrule=1pt,
    top=6pt,
    bottom=6pt,
    left=6pt,
    right=6pt
  }
}

\newtcolorbox{researchbox}[1][]{researchstyle, #1}

\begin{document}

\date{}

\title{\Large \bf From CVE Entries to Verifiable Exploits: \\An Automated Multi-Agent Framework for Reproducing CVEs}

\author{
{\rm Saad Ullah}\thanks{Corresponding author}\\
Boston University\\
saadu@bu.edu
\and
{\rm Praneeth}\\
{\rm Balasubramanian}\\
UC Santa Barbara\\
praneeth@ucsb.edu
\and
{\rm Wenbo Guo}\\
UC Santa Barbara\\
henrygwb@ucsb.edu
\and
{\rm Amanda Burnett}\\
Arizona State University\\
aburne22@asu.edu
\and
{\rm Hammond Pearce}\\
UNSW Sydney\\
hammond.pearce@\\unsw.edu.au\\
\and
{\rm Christopher Kruegel}\\
UC Santa Barbara\\
chris@cs.ucsb.edu
\and
{\rm Giovanni Vigna}\\
UC Santa Barbara\\
vigna@cs.ucsb.edu
\and
{\rm Gianluca Stringhini}\\
Boston University\\
gian@bu.edu
}

\maketitle

\begin{abstract}
High-quality datasets of real-world vulnerabilities and their corresponding verifiable exploits are crucial resources in software security research. Yet such resources remain scarce, as their creation demands intensive manual effort and deep security expertise. In this paper, we present \cvegenie, an automated, large language model (LLM)-based multi-agent framework designed to reproduce real-world vulnerabilities, provided in Common Vulnerabilities and Exposures (CVE) format, to enable creation of high-quality vulnerability datasets. 
Given a CVE entry as input, \cvegenie{} gathers the relevant resources of the CVE, automatically reconstructs the vulnerable environment, and (re)produces a verifiable exploit. Our systematic evaluation highlights the efficiency and robustness of \cvegenie's design and successfully reproduces approximately \totalcvesper{} (\totalcves{} of 841) CVEs published in 2024-2025, complete with their verifiable exploits, at an average cost of \avgcost{} per CVE. Our pipeline offers a robust method to generate reproducible CVE benchmarks, valuable for diverse applications such as fuzzer evaluation, vulnerability patching, and assessing AI's security capabilities.
\end{abstract}

\section{Introduction}
Tens of thousands of software vulnerabilities are discovered every year, with more than 40,000 vulnerabilities tracked by the National Vulnerability Database (NVD)~\cite{nvd_nvd_nodate} in 2025 alone. 
Many of these vulnerabilities are cataloged using the CVE format, which provides a standardized identification system to help organizations address security flaws in both software and hardware. Still, existing vulnerability detection tools, such as those listed by OWASP~\cite{owasp_source_nodate}, struggle to keep up with the increasing volume of code released each year. 
As a result, many vulnerabilities remain undiscovered in codebases for years (e.g., up to 2 years in Chromium and 7 years in OpenSSL~\cite{alexopoulos_foss_2022}) posing significant security risks.

Developing effective automated vulnerability detection techniques depends on high-quality datasets with reliable ground truth, as low-quality data can lead to unreliable evaluations~\cite{risse_topscore_2024} and to learn spurious correlations, especially in machine learning (ML) based approaches~\cite{arp_eval_2022, risse_limits_2023}.
Although public repositories, such as the NVD, include hundreds of thousands of CVEs, they often lack critical details, including vulnerable code, environment setup, working exploits, and validation steps, which are all required to effectively reproduce a vulnerability and to build ground truth datasets for research~\cite{mu_understanding_2018}.
This is because the main purpose of vulnerability advisories is to alert system administrators about software that needs to be updated, and concrete exploit code is often kept confidential for ethical reasons.
Unfortunately, generating this missing information from CVEs alone is labor-intensive and requires significant expertise, e.g., Mu \emph{et al.}~\cite{mu_understanding_2018} spent over 3,600 hours with 43 security expertise to reproduce only 368 memory vulnerabilities in Linux.

\begin{figure}[t]
    \centering
    \includegraphics[width=\linewidth]{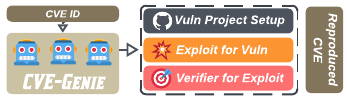}
    \caption{\cvegenie Overview.}
    \label{fig:overview}
\end{figure}

To address these challenges, the security community has adopted several approaches: manual annotation~\cite{he_sven_2023, zhou_devign_2019}, inserting bugs into real-world code~\cite{nist_sard_nodate, mirsky_vulchecker_2023}, and heuristic-based mining~\cite{ding_vulnerability_2024, chen_diversevul_2023, nikitopoulos_crossvul_2021, fan_bigvul_2020, bhandari_cvefixes_2021}.
While these methods have advanced the field, they come with limitations, such as limited support for software and Common Weakness Enumerations (CWE) categories, mislabeled vulnerabilities~\cite{risse_eval_2024}, and data becoming stale. 
For instance, the DARPA Cyber Grand Challenge dataset from 2016~\cite{darpa_cgc_nodate} quickly became outdated due to its small size and low-complexity bugs~\cite{hazimeh_magma_2020}.
Meanwhile, automated data collection helps address some of these limitations, but often tends to be limited in scope, e.g., ARVO~\cite{mei_arvo_2024} automatically curates reproducible builds with triggering inputs from OSS-Fuzz reports, but its scope is limited to memory corruption bugs in C/C++.
\textbf{This highlights a clear need for a dataset curation method that enables fully reproducible vulnerability instances across a wide range of software and vulnerability types}.

Recently, LLMs have demonstrated remarkable capabilities in solving complex software engineering (SWE) and security tasks, ranging from automated bug detection~\cite{nguyen_vuln_2024, fu_chatgptvuln_2023} and repair~\cite{pearce_examining_2023, xia_automated_2023, jiang_impact_2023}, to writing fuzzer test harnesses~\cite{deng_large_2023}. 
Moreover, the shift to agentic workflows~\cite{zhang_ai4ossfuzz_2024, chen_coder_2024, zan_java_2024, ma_understand_2024}, in which LLM-powered agents are provided access to real-world tools~\cite{abdelaziz_funccall_2024}, has further pushed the capabilities of general-purpose LLMs and led to significant performance gains on extremely difficult real-world SWE-related benchmarks, such as SWE-bench~\cite{jimenez_swebench_2024}.
This raises the question of whether LLM agents could also be useful in automatically setting up and reproducing existing CVEs, allowing us to build reliable, large, and realistic benchmarks for the security community.

In this paper, we present \cvegenie, a fully-automated, LLM-driven, multi-agent framework for end-to-end CVE reproduction. \cvegenie{} instantiates what we define as key properties of an ideal CVE reproduction system, captured by the acronym \textbf{\eager}. Specifically, an ideal system: Generates working \underline{E}xploit/proof-of-concept (PoC); Builds \underline{A}ssessors/verifiers for the exploit; \underline{G}eneralizes across CWEs, languages, and project types; Is \underline{E}nd-to-end automated; and \underline{R}ebuilds vulnerable environments.
Specifically, this paper makes the following contributions:

\begin{compactenum}
    \item We develop \cvegenie{}, an automated end-to-end CVE reproduction pipeline that extracts CVE data (e.g., software vulnerable version source code, advisories, patches, cwes, etc.), rebuilds the vulnerable environment, and generates exploits and verifiers to assess the reproduction.
    \item We design \cvegenie{} to demonstrate \emph{compositional intelligence} by a novel architecture, systematically selected LLMs and prompts, and reliable \eager-style CVE reproduction. As a result, \cvegenie{} achieves strong performance across diverse CVEs, vulnerability types, programming languages, and projects, even when CVE information is incomplete.
    \item We systematically evaluate \cvegenie{}'s architecture and demonstrate that each component is essential for optimal performance. Notably, even advanced LLMs like \texttt{o3}, when used standalone, are incapable of performing \eager-style reproduction for even a single CVE.
    \item We ran \cvegenie{} on 841 CVEs (published between June 2024 and May 2025) and successfully reproduced \totalcves{} CVEs across \totalprojects{} projects, \totalcwes{} CWEs, and \totallangs{} programming languages, highlighting our generalizability.
    \item We make our framework, source code, datasets of reproduced CVEs, and full logs of agent interactions publicly available, providing an ongoing valuable contribution for downstream research tasks (see Appendix~\ref{sec:open-science} and ~\ref{sec:applications}).
\end{compactenum}

\section{Background and Related Work}

\newcommand{\vh}[1]{\rotatebox{90}{\footnotesize\strut #1}}

\begin{table}[t]
  \footnotesize
  \setlength{\tabcolsep}{2pt}       
  \renewcommand{\arraystretch}{0.9}   
  \centering
  \begin{tabularx}{\columnwidth}{ll*{8}{c}}
    \textbf{Name} & \textbf{Source} &
    \vh{Real} & \vh{\# Projects} & \vh{\# CWEs} & \vh{\# Lang} &
    \vh{Automated} & \vh{Reproducible} & \vh{PoC} & \vh{Verifier} \\
    \midrule
    SARD~\cite{nist_sard_nodate} & Manual & \fullcircle & \textasciitilde120 & \textasciitilde150 & 6 & \redcross & \greentick & \greentick & \redcross \\
    SVEN~\cite{he_sven_2023} & Manual & \emptycircle & n/a & 9 & 3 & \redcross & \redcross & \redcross & \redcross \\
    Devign~\cite{zhou_devign_2019} & Manual & \emptycircle & 4 & n/a & 2 & \redcross & \redcross & \redcross & \redcross \\
    SecLLMHolmes~\cite{ullah_llms_2024} & Manual & \halfcircle & 4 & 8 & 3 & \redcross & \redcross & \redcross & \redcross \\
    VulChecker~\cite{mirsky_vulchecker_2023} & Manual & \fullcircle & n/a & 5 & 2 & \greentick & \redcross & \redcross & \redcross \\
    Draper~\cite{russell_draper_2018} & SA tool & \halfcircle & n/a & 9 & 2 & \greentick & \redcross & \redcross & \redcross \\
    D2A~\cite{zheng_d2a_2021} & SA tool & \emptycircle & 6 & 7 & 2 & \greentick & \redcross & \redcross & \redcross \\
    VUDENC~\cite{wartschinski_vudenc_2022} & Sec. issue & \emptycircle & 812 & 7 & 1 & \greentick & \redcross & \redcross & \redcross \\
    DiverseVul~\cite{chen_diversevul_2023} & Sec. issue & \emptycircle & 797 & 150 & 12 & \greentick & \redcross & \redcross & \redcross \\
    PrimeVul~\cite{ding_vulnerability_2024} & Patch diff & \emptycircle & 755 & 140 & 2 & \greentick & \redcross & \redcross & \redcross \\
    BigVul~\cite{fan_bigvul_2020} & Patch diff & \emptycircle & 348 & 91 & 2 & \greentick & \redcross & \redcross & \redcross \\
    CrossVul~\cite{nikitopoulos_crossvul_2021} & Patch diff & \emptycircle & 1,675 & 168 & 40 & \greentick & \redcross & \redcross & \redcross \\
    CVEfixes~\cite{bhandari_cvefixes_2021} & Patch diff & \emptycircle & 1{,}754 & 180 & 30 & \greentick & \redcross & \redcross & \redcross \\
    ARVO~\cite{mei_arvo_2024} & OSS-Fuzz & \emptycircle & 273 & 4 & 2 & \greentick & \greentick & \greentick & \redcross \\
    CVE-Bench~\cite{zhu_cvebench_2025} & Public CVEs & \emptycircle & 36 & 8 & 6 & \redcross & \greentick & \greentick & \greentick \\
    Mu \emph{et al.}~\cite{mu_understanding_2018} & Public CVEs & \emptycircle & 1 & 4 & 1 & \redcross & \greentick & \greentick & \greentick \\
    \midrule
    \cvegenie{} & Public CVEs & \emptycircle & \totalprojects{} & \totalcwes{} & \totallangs{} & \greentick & \greentick & \greentick & \greentick \\
    \bottomrule
  \end{tabularx}
  \caption{Overview of vulnerability benchmark datasets. Real: synthetic \fullcircle, real-world \emptycircle, mixed \halfcircle. \cvegenie{} results reflect CVEs from Jun~2024–May~2025.}
  \label{tab:datasets}
\end{table}

\subsection{Existing Benchmarks and Limitations}
To evaluate various vulnerability detection techniques (rule-based methods~\cite{owasp_source_nodate, yamaguchi_joern_2014}, ML-approaches~\cite{mirsky_vulchecker_2023, li_vuldeepecker_2018, lin_poster_2017, li_sysevr_2022}, and LLM-based systems~\cite{guo_unixcoder_2022, hanif_vulberta_2022, phan_cotext_2021,guo2025frontier}), the community needs standardized benchmarks.
Prior work has explored four main approaches for addressing these challenges.
(1) \emph{Manual efforts} involve analyzing~\cite{he_sven_2023} or reproducing~\cite{zhu_cvebench_2025} vulnerabilities in real-world code, or crafting synthetic examples~\cite{ullah_llms_2024}, but datasets are small, lack ecological validity, and often suffer labeling errors due to human mistakes (e.g., 50\% data in Devign~\cite{zhou_devign_2019} have wrong labels ~\cite{risse_eval_2024}).
(2) Automated labeling using \emph{static analysis (SA) tools}, such as bandit~\cite{bandit_nodate} or infer~\cite{infer_nodate}, reduces manual effort but introduces many false positives~\cite{rahman_secsmells_2019, zheng_d2a_2021}.
(3) Mining \emph{real-world data}, such as \emph{GitHub issues} or CVE \emph{patch commits}, assumes that all removed or modified code is vulnerable; However, this assumption often leads to mislabeled samples, as Risse \emph{et al.}~\cite{risse_eval_2024} highlighted for DiverseVul~\cite{chen_diversevul_2023} and BigVul~\cite{fan_bigvul_2020}. 
(4) Relying on \emph{proven sources}, such as \emph{OSS-Fuzz}, offers verified exploits and sanitizer feedback; However, these are restricted to certain bug types, primarily memory corruption in C/C++, and do not generalize across languages or vulnerability classes~\cite{mei_arvo_2024}. 

Besides data quality, existing datasets also lack necessary artifacts for dynamic evaluation beyond static labels~\cite{arp_eval_2022, risse_limits_2023,ullah_llms_2024}, including dynamic execution environments and reproducible exploits, which is difficult to obtain~\cite{mu_understanding_2018}.
Table \ref{tab:datasets} provides an overview of the current state of vulnerability benchmarks.

\subsection{Call for \eager{} Style for CVE Reproduction}
A \emph{CVE ID} is a unique identifier for publicly disclosed security vulnerabilities in real-world software. Each CVE ID corresponds to a \emph{CVE entry} that includes key details for that CVE, such as vulnerability description, source code, affected versions, security advisories, CWE classifications, and patches, enabling security professionals to assess and mitigate risks. Creating a dataset from the standard CVE database is a promising way towards addressing the aforementioned challenges (see Appendix~\ref{sec:applications}). Therefore, we introduce \eager{}, a set of crucial criteria for ideal CVE reproduction:

\begin{compactitem}
    \item \underline{\textbf{E}}xploit Generation -- (re)creation of an exploit or PoC that reliably triggers the vulnerability in the CVE.
    \item \underline{\textbf{A}}ssessment -- inclusion of a ``verifier'' or ``sanitizer'' capable of assessing whether the generated exploit or PoC successfully triggers the vulnerability.
    \item \underline{\textbf{G}}eneralization -- reproduction of CVEs across diverse CWEs, programming languages, and software projects.
    \item \underline{\textbf{E}}nd-to-end Automation -- execution of all stages of CVE reproduction in a fully automated manner.
    \item \underline{\textbf{R}}ebuild project -- reconstruction of the original vulnerable environment to facilitate exploit/PoC execution.
\end{compactitem}
However, none of the existing methods satisfy all the criteria (see Table~\ref{tab:techniques}).
For example, manual effort for producing \eager{} datasets does not scale and is not generalizable~\cite{mu_understanding_2018,zhu_cvebench_2025}.
ARVO~\cite{mei_arvo_2024} focuses solely on memory-related bugs from OSS-Fuzz C/C++ projects, limiting its scope.
In contrast, \cvegenie{} is the only framework that satisfies all criteria of \eager{} reproduction.

\begin{table}[t]
    \small
    \centering
    \begin{tabularx}{0.98\linewidth}{l*{5}{>{\centering\arraybackslash}X}}
        \toprule
        \textbf{Method} & \textbf{\underline{E}xploit} & \textbf{\underline{A}ssess} & \textbf{\underline{G}eneral} & \textbf{\underline{E}2E} & \textbf{\underline{R}ebuild} \\
        \midrule
        Manual~\cite{ma_understand_2024, zhu_cvebench_2025} & \greentick & \greentick & \redcross & \redcross & \greentick \\
        Fuzzers & \greentick & \redcross & \redcross & \redcross & \redcross \\
        Metasploit~\cite{metasploit_nodate} & \greentick & \redcross & \redcross & \redcross & \redcross \\
        ARVO~\cite{mei_arvo_2024} & \greentick & \redcross & \redcross & \greentick & \greentick \\
        \cvegenie & \greentick & \greentick & \greentick & \greentick & \greentick \\
        \bottomrule
    \end{tabularx}
    \caption{Comparing Potential Vulnerability Reproduction Methods and their \eager{} attributes.}
    \label{tab:techniques}
\end{table}

\subsection{Large Language Models and Agents}
LLMs are transformer-based neural network models with billions of parameters. 
They have demonstrated remarkable performance across various application domains, including question answering~\cite{openai_gpt_2024}, mathematical reasoning~\cite{ahn_math_2024}, and code generation~\cite{chen_evaluating_2021, austin_program_2021}. 
Recent work also shows their utility in security tasks, e.g., bug detection and repair~\cite{pearce_examining_2023, xia_automated_2023, jiang_impact_2023} and writing fuzzer test harnesses~\cite{deng_large_2023}.

LLM agents are systems that combine LLMs with tools.
They can finish complex tasks via a sequence of actions, including planning, generation, and tool executions~\cite{talebirad_multi-agent_2023, he_llm-based_2025, park_generative_2023, li_camel_2023, jin_llms_2024}. 
An agentic system can have multiple agents, each with a sub-task, such as reasoning, critiquing, generating, perceiving, remembering, or teaching, as well as its own workflow and tool sets.
In security, researchers have started to construct LLM agents to find and repair vulnerabilities.
There are multi-agent patching systems, where different sub-agents are responsible for fault localization, patch generation, and validation (e.g., PatchPilot~\cite{li2025patchpilot}, and PatchAgent~\cite{yupatchagent}). 
Through the recent DARPA AIxCC competition, researchers constructed agentic-based systems with end-to-end capabilities for vulnerability detection and patching~\cite{guo2025frontier}.
These recent successes provide the argument for using LLM agents in vulnerability analysis and software development, as well as for building an \eager-style automated end-to-end CVE reproduction framework.

\section{\cvegenie}

\begin{figure*}[t]
    \centering
    \includegraphics[width=\linewidth]{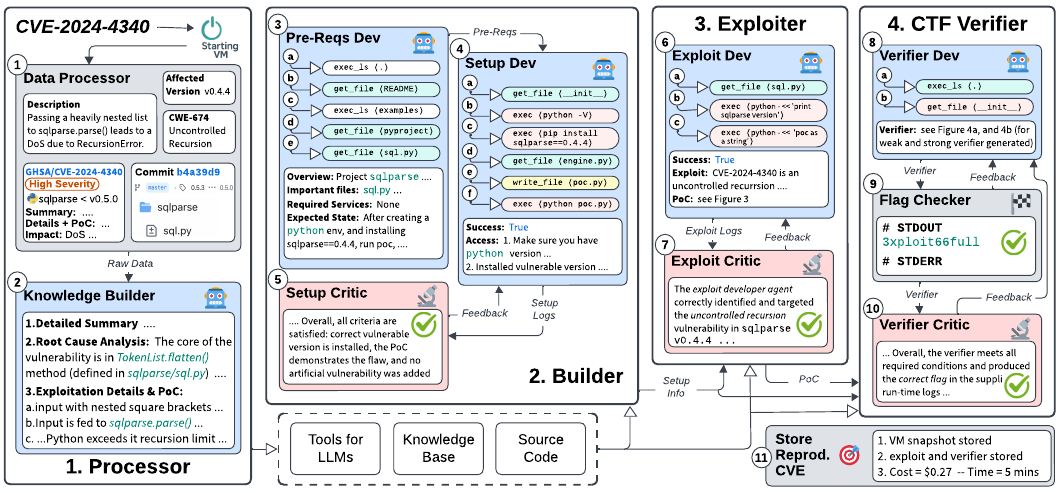}
    \caption{\cvegenie{} architecture and an end-to-end example of workflow of reproduction for CVE-2024-4340, i.e., Denial of Service due to \texttt{RecursionError} in \texttt{sqlparse < v0.5.0}. See artifact of CVE-2024-4340 complete reproduction run here - \url{https://github.com/BUseclab/cve-genie/tree/main/results/CVE-2024-4340}}.
    \label{fig:architecture}
\end{figure*}

We design \cvegenie{} to exhibit \emph{compositional intelligence}, i.e., the ability to solve complex tasks by decomposing them into meaningful sub-tasks whose outputs are explicitly composed and verified to form a correct end-to-end solution. To this end, first, we design a \ul{novel architecture} that decomposes CVE understanding and reproduction into four interdependent modules: (1) the \cveprocessor{} (Section~\ref{subsec:cve_processor}), which retrieves source code and constructs a structured knowledge base; (2) the \project{} (Section~\ref{subsec:project_builder}), which reconstructs the vulnerable environment; (3) the \exploit{} (Section~\ref{subsec:exploit_builder}), which reproduces the exploit; and (4) the \verifier{} (Section~\ref{subsec:verifier-builder}), which generates a verifier for the exploit. 
Second, rather than relying on a single LLM or a fixed prompting strategy across tasks, we \ul{systematically select the most suitable LLMs and engineer task-specific prompts for each component} of \cvegenie{}, resulting in an optimized end-to-end configuration (Section~\ref{subsec:llms-cve-genie}).
Third, we bridge the gap identified in Section~\ref{tab:techniques}, namely the lack of any system capable of \ul{true \textsc{EAGER}-style CVE reproduction}, and ground \cvegenie's design in the following guiding principles:

\begin{compactenum}
    \item \textbf{Modular Task Decomposition:} LLMs struggle with long, complex contexts~\cite{ullah_llms_2024} and tasks~\cite{jimenez_swebench_2024}. Thus, \cvegenie{} decomposes reproduction into focused modules and sub-modules handled by specialized agents with systematically engineered prompts, providing the most suitable architecture for end-to-end CVE reproduction (Section~\ref{subsec:design-ablation}).

    \item \textbf{Robustness to Incomplete Data:} Since missing CVE fields can lower reproduction success by up to 44\%~\cite{mu_understanding_2018}, \cvegenie{} mitigates this by relying on patch or source-code analysis when advisories/PoCs are unavailable (Section~\ref{subsec:robust}).

    \item \textbf{Reliability through Self-Critique:} To counter LLMs’ reasoning limits~\cite{ullah_llms_2024}, each module uses paired \emph{developer} and \emph{critic} agents in a ReAct-style loop~\cite{yao_react_2022}, enabling iterative refinement and internal verification (Section~\ref{subsec:design-ablation}).
\end{compactenum}
Together, these principles elevate \cvegenie{} beyond a fixed engineering pipeline and enable \cvegenie{} to operationalize \emph{compositional intelligence} in a fully automated setting, yielding the first practical, generalizable, and end-to-end framework for \eager-style CVE reproduction (see Appendix~\ref{app:tools} for further details).
The following subsections detail the workings of each \cvegenie{} component, using CVE-2024-4340 as a running example corresponding to the end-to-end reproduction shown in Figure~\ref{fig:architecture}. Moreover, we provide a detailed case study on CVE-2024-5129, a logic vulnerability of missing authorization in \verb|lunary-ai|, in Appendix~\ref{sec:case-study-cve-2024-5129}.

\subsection{Processor}
\label{subsec:cve_processor}
This module comprises two sub-modules: the \dataprocessor, which locates the vulnerable project's source code and gathers relevant raw resources, and the LLM-based \knowledge, which transforms these resources into a structured knowledge base. Below, we provide detailed functionalities of these sub-modules.

\subsubsection{Data Processor \cirnum{1}}
\label{subsubsec:data-processor}

For the given CVE ID, this sub-module identifies and extracts the vulnerable version of the software along with all relevant CVE details, as follows:

\descrit{Source Code Extraction.} The \dataprocessor{} begins by locating the \ul{source code} for the vulnerable project linked to the CVE. It searches the GitHub URLs referenced in the ``cvelist'' \footnote{``\emph{cvelist}'' is an automated pilot program for CVE submission through GitHub  (https://github.com/CVEProject/cvelist)} to identify the project repository. For closed-source software, \emph{Data Processor} allows users to supply the repository URL. Once the source code is located, it identifies \ul{affected software configurations}: vulnerable versions, platforms, or settings. For instance, CVE-2024-4340 affects all versions of ``sqlparse'' prior to \verb|v0.5.0|. Accordingly, the latest affected version (\verb|v0.4.4|) is retrieved using the GitHub API, and its \emph{source code} is downloaded (as shown in Figure~\ref{fig:architecture}).

\descrit{Vulnerability Information Extraction.} After the vulnerable version of the project source code is downloaded, \dataprocessor{} extracts the following four key pieces of information from ``cvelist'', if available: 
(1) \emph{CVE description}: A high-level summary of the vulnerability in the target project. 
(2) \emph{CWE data}: CWE information, a categorization system for hardware and software weaknesses that associates the CVE with a specific vulnerability type, e.g., CWE-674 (Uncontrolled Recursion).
(3) \emph{Patch commits}: The \texttt{git-diff} of code changes in the target source code that fix the vulnerability. These commits help LLMs localize the issue, understand its root cause~\cite{he_sven_2023, ding_vulnerability_2024, fan_bigvul_2020}, and generate effective exploits. 
(4) \emph{Security advisories and PoC}: We filter URLs referenced in the CVE entry by keywords (e.g., ``security'', ``advisory'', ``bounty'', etc.), and scrape their contents.

\descr{Output.} The vulnerable version of a project's source code alongside its directory structure and CVE raw data.

\subsubsection{Knowledge Builder \texorpdfstring{\cirnum{2}}{}}

This agent is instructed to analyze, distill, and organize the raw information, extracted by the \dataprocessor, into a structured knowledge base while retaining essential details required to reproduce the given CVE. This includes extracting PoC or exploit instructions from security advisories, which are invaluable for accurately reproducing the vulnerability. For instance, CVE-2024-4340 has a PoC provided in its security advisory, and the patch commit helps understand the root cause by highlighting where/how the vulnerability was mitigated. As shown in Figure~\ref{fig:architecture}, the \knowledge{} included both of these details in the knowledge base. In case an advisory does not include an exploit, the LLM agent is tasked with generating an overview of what an exploit might entail. Additionally, capturing the patch details helps to localize the root cause of the vulnerability~\cite{he_sven_2023, ding_vulnerability_2024, bhandari_cvefixes_2021} and aids in crafting an effective exploit. This knowledge base acts as the long-term memory for agents, helping them to effectively reproduce CVEs, by including only essential information in their context to avoid overloading the LLM's context window.

\descr{Output.} A structured \emph{CVE knowledge base} providing details of CVE, including its associated CWEs, affected software configurations, root causes (if provided in the security advisory), and essential information from security advisories and patch commits.

\subsection{Builder}
\label{subsec:project_builder}
Once the knowledge base is populated and the vulnerable version of the project's source code is downloaded, the next objective is to build the project in a way that allows the exploit to be executed. To achieve this, our pipeline employs two \emph{developer} LLM agents: the \emph{\predev{} Agent}, which analyzes and plans the environment setup, and the \emph{\setupdev{} Agent}, which executes this plan to configure the vulnerable environment. As well as one \emph{critic}, the \emph{\setupcritic{} Agent}, which analyzes the logs of the \emph{\setupdev{}} and evaluates the setup for the project, allowing the system to correct its mistakes. Below, we discuss these in detail.

\subsubsection{Prerequisite Developer Agent \texorpdfstring{\cirnum{3}}{}}

We introduce this agent as an initial exploration and planning step before actually setting up the vulnerable project. This is done for three key reasons.\\
(1) Our preliminary experiments reveal that many projects contain inaccurate and incomplete information in their \verb|README| files (e.g., vulnerable version \texttt{v1.2.7} of \texttt{lunary} for CVE-2024-5129, includes incorrect paths to \texttt{.env} files that are crucial for the setup). Without the \emph{Prerequisite agent}, these details were consistently overlooked by LLMs, causing setup failures. Identifying and correcting such issues early reduces unsuccessful attempts, conserves the LLM's context window, and improves the likelihood of successful setup.\\
(2) During environment setup, the agent's job is to explore and analyze the source code to identify critical components and to build the project by executing a series of commands. For large projects, having to do both tasks with one agent can exceed the context limit and reduce efficiency (as demonstrated in the ablation study in Section~\ref{subsec:design-ablation}). To address this, we delegate the exploration of the code base to a dedicated agent, the \predev, allowing the setup process to focus more on building the project.\\
(3) The \predev{} also defines the ``expected state,'' i.e., when the project is fully set up and ready for exploitation. This state guides the \emph{Setup Developer Agent} in verifying the correct configuration of the vulnerable environment. For example, in CVE-2024-4340 (Figure~\ref{fig:architecture}), the \predev{} explored the root directory and key files such as \texttt{README} and \texttt{sql.py} to understand the project context, then specified the expected state as having \texttt{sqlparse} version \texttt{v0.4.4} installed for the exploit to work correctly.

\descr{Tools.} For this step, we only provide the LLM agent with access to read-only tools, i.e., \verb|execute_ls_command|, and \verb|get_file_by_name|, which the agent can use to traverse the code base and read files, because we do not want the agent to execute any commands or write to files in this phase (see Appendix~\ref{app:tools} for tools details).

\descr{Output.} (1) Detailed \emph{overview} of the project, (2) \emph{important files} to pay attention to during the setup, (3) \emph{required services} and their configurations, and (4) \emph{expected state} of the project upon successful setup to facilitate verification.

\subsubsection{Setup Developer Agent \texorpdfstring{\cirnum{4}}{}}

The \setupdev{} begins in the directory containing the vulnerable version’s source code and focuses on configuring the project based on instructions from the \predev{}, exploring the codebase when necessary. For CVEs involving third-party libraries, our initial experiments showed that setup is greatly simplified by using package managers like \texttt{pip} or \texttt{npm} to install specific vulnerable versions instead of building from source. For example, in the case of CVE-2024-4340 affecting \texttt{sqlparse}, the \predev{} instructed the installation of version \texttt{0.4.4}, and the \setupdev{} executed \verb|pip install sqlparse==0.4.4| directly (command \texttt{4a}, Figure~\ref{fig:architecture}), resulting in a smooth setup. If the package manager fails, the \setupdev{} falls back to building from source. Typically, the \predev{} also recommends running a basic PoC to confirm readiness before handing off to the next agent (e.g., commands \texttt{4e} and \texttt{4f} in Figure~\ref{fig:architecture}).

\descr{Tools.} For this step, we allow access to the following tools, i.e., \verb|execute_ls_command|, \verb|execute_linux_command|, \verb|get_file|,
\verb|write_to_file|, and \verb|set_environment_variable|. 

\descr{Output.} This agent returns a \emph{final decision} whether the setup is successful or not. 
If successful, the output also includes instructions on how another agent can \emph{access} the running project.

\subsubsection{Setup Critic Agent \texorpdfstring{\cirnum{5}}{}}

The \setupcritic{} evaluates whether the project setup performed by the \setupdev{} is correct and complete, ensuring that the environment is properly configured for the vulnerability described in the CVE knowledge base to be exploited. Additionally, this agent plays a critical role in detecting deceptive or shortcut behaviors by the LLM, such as fabricating a mock project instead of performing a genuine setup, or deliberately injecting vulnerabilities into the codebase to simplify exploitation. These undesirable patterns are discussed in detail in Section~\ref{subsubsec:llm-project}, which informs the instruction design for the \setupcritic{}. Based on this, the \emph{critic} is equipped to identify both functional and security-related flaws in the setup process.

\descr{Output.} A comprehensive \emph{analysis} of the setup logs, a binary \emph{decision} on whether the setup is valid and complete, and actionable \emph{feedback} for correcting issues or improving the setup if necessary.

\begin{figure}[t]
\centering
\begin{subfigure}{\linewidth}
\centering
\lstset{
    language=Python,
    basicstyle=\footnotesize\ttfamily,
    backgroundcolor=\color{gray!10},
    numbers=left,
    numberstyle=\tiny\color{gray}\ttfamily,
    stepnumber=1,
    numbersep=5pt,
    showspaces=false,
    showstringspaces=false,
    showtabs=false,
    frame=none,
    framesep=0pt,
    tabsize=4,
    breaklines=true,
    breakatwhitespace=false,
    columns=fullflexible,
    keepspaces=true,
    xleftmargin=0pt,
    xrightmargin=0pt,
    aboveskip=0pt,
    belowskip=0pt,
    lineskip=0pt,
    keywordstyle=\color{green!50!black}\bfseries,
    commentstyle=\color{red!70!black}\itshape,
    stringstyle=\color{red!70!black},
    emphstyle=\color{blue!80!black}\bfseries,
    emph={ValueError}
}
\setlength{\fboxsep}{0pt}
\begin{lstlisting}[escapechar=@]
"""
PoC - CVE-2024-4340 (sqlparse < 0.5.0)
EXAMPLE THAT CRASHES v0.4.4
@\colorbox{cyan!17}{\strut\hspace{-\fboxsep}\makebox[\linewidth][l]{\textcolor{red!70!black}{\texttt{~~~~}python3 exploit.py 10000}}}@
"""
import sys
import sqlparse
def main() -> None:
    if len(sys.argv) != 2:
        print("Usage: python3 exploit.py <depth>")
        sys.exit(1)
    arg = sys.argv[1]
    try:
        depth = int(arg)
@\colorbox{cyan!17}{\strut\hspace{-\fboxsep}\makebox[\linewidth][l]{\textcolor{black}{\texttt{~~~~~~~~payload~=~"["~*~depth~+~"]"~*~depth}}}}@
    except ValueError:
        payload = arg
@\colorbox{cyan!17}{\strut\hspace{-\fboxsep}\makebox[\linewidth][l]{\textcolor{black}{\texttt{~~~~}}sqlparse.parse(payload)}}@
if __name__ == "__main__":
    main()
\end{lstlisting}
\caption{Exploit script for CVE-2024-4340. It constructs a deeply nested list payload and submits it to \texttt{sqlparse.parse()}, triggering uncontrolled recursion. The included comment documents a crashing input (depth 10,000) that reliably triggers the vulnerability.}
\label{subfig:exploit-script}
\end{subfigure}

\hfill

\begin{subfigure}{\linewidth}
\centering
\lstset{
    language=Python,
    basicstyle=\footnotesize\ttfamily,
    backgroundcolor=\color{red!4},
    numbers=left,
    numberstyle=\tiny\color{gray}\ttfamily,
    stepnumber=1,
    numbersep=5pt,
    showspaces=false,
    showstringspaces=false,
    showtabs=false,
    frame=none,
    framesep=0pt,
    tabsize=4,
    breaklines=true,
    breakatwhitespace=false,
    columns=fullflexible,
    keepspaces=true,
    xleftmargin=0pt,
    xrightmargin=0pt,
    aboveskip=0pt,
    belowskip=0pt,
    lineskip=0pt,
    keywordstyle=\color{blue!80!black}\bfseries,
    commentstyle=\color{green!60!black},
    stringstyle=\color{red!70!black},
    emphstyle=\color{red!70!black}\bfseries,
    emph={RecursionError, yield, from, in},
}
\setlength{\fboxsep}{0pt}
\begin{lstlisting}[escapechar=@]
@\colorbox{red!20}{\strut\hspace{-\fboxsep}\makebox[\linewidth][l]{\textcolor{black}{\texttt{Traceback~(most~recent~call~last):}}}}@
  File "poc.py", line 20, in <module>
    sqlparse.parse(payload)
  ....
  "sqlparse/sql.py", line 214, in flatten
    yield from token.flatten()
  [Previous~line~repeated~983~more~times]
@\colorbox{red!20}{\strut\hspace{-\fboxsep}\makebox[\linewidth][l]{\textcolor{red!70!black}{\bfseries\texttt{RecursionError}}\textcolor{black}{\texttt{:~maximum~recursion~depth~exceeded}}}}@
\end{lstlisting}
\caption{Execution traceback produced by the exploit, showing a \texttt{RecursionError} after excessive recursive calls within \texttt{sqlparse}.}
\label{subfig:exploit-trigger}
\end{subfigure}
\caption{PoC for CVE-2024-4340 in \texttt{sqlparse v0.4.4}, showing the exploit script and the resulting crash.}
\label{fig:exp-example}
\end{figure}

\subsection{Exploiter}
\label{subsec:exploit_builder}
After successfully configuring the vulnerable system, the next step in our pipeline is to generate an exploit for the given vulnerability. \cvegenie{} accomplishes this using two LLM agents: the \exploitdev, which develops the PoC for an exploit in the vulnerable environment, and the \exploitcritic, which analyzes the logs of the \exploitdev{} and evaluates the exploit for the given vulnerability. Below, we discuss their detailed functionalities.

\subsubsection{Exploit Developer Agent \texorpdfstring{\cirnum{6}}{}}
\label{subsubsec:exp-dev}

The \exploitdev{} is responsible for crafting and demonstrating exploits for vulnerabilities listed in the CVE knowledge base on pre-configured vulnerable systems. If a PoC is provided, the agent replicates and verifies it by executing the script with the appropriate triggering input before delivering the final, working PoC. For example, in CVE-2024-4340 (Figure~\ref{fig:architecture}), the PoC was available, so the agent validated the setup (\texttt{6b}), demonstrated the exploit (\texttt{6c}) by triggering \texttt{RecursionError} in \texttt{sqlparse/sql.py} (see Figure~\ref{subfig:exploit-trigger}), and submitted the verified PoC script (see Figure~\ref{subfig:exploit-script}).
If no PoC or exploitation steps are available, the \exploitdev{} iteratively analyzes the codebase using available tools to develop a functional exploit. Upon success, it produces a Python PoC script that accepts the crashing input via the command line, includes comments on the expected input format, and provides an example input. This ensures that the \verifier{} can reliably validate the exploit without repeating the analysis.

\descr{Tools.} This agent uses the same tools as the \setupdev{}.

\descr{Output.} The agent returns its \emph{final decision} whether it considers the exploit is successful or not.
If successful, it provides an \emph{overview of the exploit} and a \texttt{Python} PoC script.

\subsubsection{Exploit Critic Agent \texorpdfstring{\cirnum{7}}{}}
\label{subsubsec:exploit-critic}

Similar to the role of the \setupcritic{}, the \exploitcritic{} is tasked with evaluating the behavior of the \exploitdev{} by analyzing its execution logs. Its primary goal is to determine whether the generated exploit is both valid and high-quality. This includes verifying that the exploit could plausibly succeed in a real-world setting; that it does not rely on manipulating or altering the original setup in illegitimate ways; and that it avoids fabricated shortcuts, oversimplified logic, or artificial assumptions that would not hold in practice.

To perform this assessment, the \exploitcritic{} leverages a predefined set of behavioral patterns and common failure cases, as detailed in Section~\ref{subsubsec:llm-exploit}. By comparing the exploit's behavior against these known issues, the critic can detect actions that are unfaithful to the original intent or technically incorrect. When flaws are identified, the \exploitcritic{} provides structured, actionable feedback to improve the robustness of the exploit in subsequent iterations.

\subsection{CTF Verifier}
\label{subsec:verifier-builder}

After generating a working exploit for a given CVE, the final step is to create a verifier script that confirms whether the exploit was successful. This approach is inspired by security Capture-The-Flag (CTF) challenges, where exploiting a vulnerability reveals a hidden ``flag'' i.e., a string of random characters or numbers. The \verifier{} uses three sub-modules to develop and validate such verifiers. The \verifierdev{} agent module generates a candidate verifier, the \flagcheck{} module ensures that the exploit correctly retrieves a flag, and the \verifiercritic{} agent assesses the quality and effectiveness of the verifier for the given exploit.

\begin{figure}[]
\centering

\begin{subfigure}{\linewidth}
\centering
\lstset{
    language=Python,
    basicstyle=\footnotesize\ttfamily,
    backgroundcolor=\color{gray!10},
    numbers=left,
    numberstyle=\tiny\color{gray}\ttfamily,
    stepnumber=1,
    numbersep=5pt,
    showspaces=false,
    showstringspaces=false,
    showtabs=false,
    frame=none,
    framesep=0pt,
    tabsize=4,
    breaklines=true,
    breakatwhitespace=false,
    columns=fullflexible,
    keepspaces=true,
    xleftmargin=0pt,
    xrightmargin=0pt,
    aboveskip=0pt,
    belowskip=0pt,
    lineskip=0pt,
    keywordstyle=\color{green!50!black}\bfseries,
    commentstyle=\color{red!70!black}\itshape,
    stringstyle=\color{red!70!black},
}
\setlength{\fboxsep}{0pt}
\begin{lstlisting}[escapechar=@]
import os, subprocess, sys
FLAG = "3xploit66full"

def pre_setup():
@\colorbox{red!15}{\strut\hspace{-\fboxsep}\makebox[\linewidth][l]{\textcolor{red!70!black}{\texttt{~~~~\#~Configuring~PYTHONPATH}}}}@

def run_exploit():
    cmd = [sys.executable, "exploit.py", "10000"]
@\colorbox{red!15}{\strut\hspace{-\fboxsep}\makebox[\linewidth][l]{\texttt{~~~~}\textcolor{green!50!black}{\bfseries\texttt{return}}\texttt{~}subprocess.run(cmd,~timeout=20)}}@

def post_exploit(proc):
    output = (proc.stdout or "") + (proc.stderr or "")
@\colorbox{red!15}{\strut\hspace{-\fboxsep}\makebox[\linewidth][l]{\texttt{~~~~}\textcolor{green!50!black}{\bfseries\texttt{return}}\texttt{~proc.returncode~!=~0~}\textcolor{green!50!black}{\bfseries\texttt{and}}}}@
@\colorbox{red!15}{\strut\hspace{-\fboxsep}\makebox[\linewidth][l]{\texttt{~~~~~~~~~~~~}\textcolor{red!70!black}{\texttt{"RecursionError"}}\texttt{~}\textcolor{green!50!black}{\bfseries\texttt{in}}\texttt{~output}}}@

if __name__ == "__main__":
    pre_setup()
    try: result = run_exploit()
@\colorbox{red!15}{\strut\hspace{-\fboxsep}\makebox[\linewidth][l]{\texttt{~~~~}\textcolor{green!50!black}{\bfseries\texttt{except}}\texttt{~}subprocess.TimeoutExpired:~sys.exit(1)}}@
    if post_exploit(result):
        print(FLAG)
\end{lstlisting}
\caption{Attempt 1 -- Weak verifier script: \texttt{pre\_setup()} lacks a check for the vulnerable \texttt{sqlparse} version. The exploit is run with a \texttt{20s} timeout, risking missed \texttt{RecursionErrors} if they occur after the limit. The verification relies solely on detecting the string \texttt{RecursionError} in output, which can be trivially spoofed.}
\label{subfig:verifier-inc}
\end{subfigure}

\hfill

\begin{subfigure}{\linewidth}
\centering
\lstset{
    language=Python,
    basicstyle=\footnotesize\ttfamily,
    backgroundcolor=\color{gray!10},
    numbers=left,
    numberstyle=\tiny\color{gray}\ttfamily,
    stepnumber=1,
    numbersep=5pt,
    showspaces=false,
    showstringspaces=false,
    showtabs=false,
    frame=none,
    framesep=0pt,
    tabsize=4,
    breaklines=true,
    breakatwhitespace=false,
    columns=fullflexible,
    keepspaces=true,
    xleftmargin=0pt,
    xrightmargin=0pt,
    aboveskip=0pt,
    belowskip=0pt,
    lineskip=0pt,
    keywordstyle=\color{green!50!black}\bfseries,
    commentstyle=\color{red!70!black}\itshape,
    stringstyle=\color{red!70!black},
}
\setlength{\fboxsep}{0pt}
\begin{lstlisting}[escapechar=@]
import os, sys, importlib, traceback
FLAG = "3xploit66full"

def pre_setup():
@\colorbox{cyan!17}{\strut\hspace{-\fboxsep}\makebox[\linewidth][l]{\texttt{~~~~}\textcolor{green!50!black}{\bfseries\texttt{assert}}\texttt{~sqlparse.\_\_version\_\_~<~}\textcolor{red!70!black}{\texttt{"0.5.0"}}}}@

def run_exploit_inprocess():
    sys.argv = ["exploit.py", "10000"]
    try:
@\colorbox{cyan!17}{\strut\hspace{-\fboxsep}\makebox[\linewidth][l]{\texttt{~~~~~~~~}importlib.import\_module(\textcolor{red!70!black}{\texttt{"exploit"}}).main()}}@
        return "no_exception", ""
@\colorbox{cyan!17}{\strut\hspace{-\fboxsep}\makebox[\linewidth][l]{\texttt{~~~~}\textcolor{green!50!black}{\bfseries\texttt{except}}\textcolor{red!70!black}{\bfseries\texttt{~RecursionError:~}}}}@
@\colorbox{cyan!17}{\strut\hspace{-\fboxsep}\makebox[\linewidth][l]{\texttt{~~~~~~~~}\textcolor{green!50!black}{\bfseries\texttt{return}}\texttt{~}\textcolor{red!70!black}{\texttt{"recursion\_error"}},~traceback.format\_exc()}}@
    except: 
        return "other_exception", traceback.format_exc()

def exploit_succeeded(status, tb):
@\colorbox{cyan!17}{\strut\hspace{-\fboxsep}\makebox[\linewidth][l]{\texttt{~~~~}\textcolor{green!50!black}{\bfseries\texttt{return}}\texttt{~status~==~}\textcolor{red!70!black}{\texttt{"recursion\_error"}}\texttt{~}\textcolor{green!50!black}{\bfseries\texttt{and}}}}@
@\colorbox{cyan!17}{\strut\hspace{-\fboxsep}\makebox[\linewidth][l]{\texttt{~~~~~~~~~~~~}\textcolor{red!70!black}{\texttt{"sqlparse/sql.py"}}\texttt{~}\textcolor{green!50!black}{\bfseries\texttt{in}}\texttt{~tb}}}@

if __name__ == "__main__":
    pre_setup()
    status, tb = run_exploit_inprocess()
    if exploit_succeeded(status, tb):
        print(FLAG)
\end{lstlisting}
\caption{Attempt 2 with \verifiercritic's feedback -- Improved verifier: Validates project version \texttt{sqlparse < 0.5.0}, runs the exploit in-process to avoid timeout issues, and explicitly checks for a genuine \texttt{RecursionError} with \texttt{traceback} originating from \texttt{sqlparse/sql.py}, preventing spoofed or misleading outputs.}
\label{subfig:verifier-crr}
\end{subfigure}

\caption{Verifier scripts for exploit (in Figure~\ref{fig:exp-example}) for CVE-2024-4340 corresponding to the run in Figure~\ref{fig:architecture}, illustrating the progression from a weak to a robust verifier based on \verifiercritic{} feedback.}
\label{fig:verifier-example}
\end{figure}

\subsubsection{Verifier Developer Agent \texorpdfstring{\cirnum{8}}{}}
\label{subsubsec:ctf-ver-dev}

Following the CTF methodology, we prompt the \verifierdev{} to create a general CTF-style \verb|Python| verifier script for a given PoC script. So, if the verifier script runs the PoC and it successfully triggers the vulnerability, it returns this flag (\texttt{3xploit66ful}). We prompt the \verifierdev{} to format the verifier script in three structured steps:

\begin{compactenum}
    \item \textit{Pre-Setup}: Prepares the necessary inputs for the exploit, sets up the environment, and prepares a flag to return if the exploit correctly triggers the vulnerability.
    \item \textit{Exploit Execution}: Runs the provided PoC script to trigger the vulnerability. We make sure that the agent cannot modify the PoC script to avoid contamination.
    \item \textit{Post-Setup}: Verifies the success of the exploit, and if successful, the script returns the flag.
\end{compactenum}
This verifier architecture provides consistent and reliable exploit validation across a broad spectrum of vulnerability classes. It supports vulnerabilities with directly observable failure signals, such as a \texttt{RecursionError} in CVE-2024-4340 or an \texttt{AddressSanitizer: heap-buffer-overflow} in CVE-2024-10525, as well as higher-level logic flaws. For example, in the case of the missing-authorization flaw in CVE-2024-5129 (Appendix \ref{sec:case-study-cve-2024-5129}), the verifier first inserts a legitimate dataset with a known UUID, then executes the exploit by issuing an unauthenticated deletion request targeting that UUID, and finally confirms successful exploitation by verifying that the dataset has been removed from the database.

\descr{Tools.} This agent follows the same read-only tools as the \predev{} agent, as in this step we don't want the model to execute anything, and we do all validations using the \flagcheck{} and the \verifiercritic{}.

\descr{Output.} A \verb|Python| script that acts as a custom \emph{verifier} for the exploit/PoC of the given CVE.

\subsubsection{Flag Checker \texorpdfstring{\cirnum{9}}{}}
\label{subsubsec:func-eval}

Once the verifier script is generated for the given exploit, before passing it to the \emph{critic} agent, the \flagcheck{} executes the verifier script to ensure that it runs with the exploit script without any errors and produces the expected output, i.e., the correct flag. For instance, figure~\ref{fig:architecture} shows that the correct flag \texttt{3xploit66ful} is released when executing the verifier scripts generated for CVE-2024-4340. If a script fails due to a runtime error or if the flag is missing from the output, the system assumes that the verifier is incorrect. Feedback is then sent to the \verifierdev{} to revise the script. This process is repeated until a maximum of five attempts is reached, with each attempt ensuring functional correctness of the verifier.%

\subsubsection{Verifier Critic Agent \texorpdfstring{\cirnum{10}}{}}
\label{subsubsec:ver-critic}

Once the verifier and exploit scripts pass functional validation via the \flagcheck{}, they are further examined by the \verifiercritic{}, which performs a similar evaluative role as the other critic agents, focusing specifically on the response produced by the \verifierdev{}. It examines whether the verification process was properly carried out, without assumptions, omissions, or unreliable heuristics. Using the behavioral issues identified in Section~\ref{subsubsec:llm-verify}, the agent flags incomplete or misleading verification strategies and provides feedback on how the verification could be made more robust or accurate. For instance, Figure~\ref{fig:verifier-example} shows a detailed example of verifiers generated for CVE-2024-4340 with weak and bypassable verification criteria, which were corrected using the \verifiercritic's feedback.

\subsubsection{Store Reproduced CVE \texorpdfstring{\cirnum{11}}{}}
\label{subsec:store}

If the given CVE passes the \verifiercritic{} check, \cvegenie{} marks it as reproduced and stores the VM snapshot for the vulnerable environment as well as the \emph{exploit} and \emph{verifier} scripts. Moreover, \cvegenie{} also stores the metadata, such as LLM cost, time spent, and all agents' conversations.

\subsection{Selecting LLMs and Engineering Prompts for \cvegenie's Optimal Performance}
\label{subsec:llms-cve-genie}

Upon finalizing the design of \cvegenie{}’s architecture, we perform a systematic LLM model and prompt selection process to identify the best performing LLMs and prompt integrations. We evaluate ten state-of-the-art LLMs (Table~\ref{tab:llms}) across \cvegenie{}’s three core modules, \project{}, \exploit{}, and \verifier{}, using a diverse set of 60 post–knowledge-cutoff CVEs, while simultaneously engineering and refining task-specific prompts for each module.

\begin{table}[t]
    \small
    \centering
    \begin{tabularx}{\linewidth}{l@{\hspace{0.3cm}}p{1cm}@{\hspace{0.2cm}}p{0.7cm}@{\hspace{0.25cm}}p{0.5cm}@{\hspace{0.25cm}}p{0.7cm}@{\hspace{0.3cm}}p{0.5cm}@{\hspace{0.3cm}}p{0.5cm}}
        \toprule
        \textbf{Model} & \textbf{\# Params} & \textbf{Max Inp} & \textbf{Max Out} & \textbf{Reas-oning} & \textbf{Open Src} & \textbf{Cutoff} \\
        \midrule
        o3 & n/a & 200k & 100k & \centering \greentick & \centering \redcross & 05/2024 \\
        o4-mini & n/a & 200k & 100k & \centering \greentick & \centering \redcross & 05/2024 \\
        Claude 3.7 Sonnet & n/a & 200k & 64k & \centering \greentick & \centering \redcross & 11/2024 \\
        Claude 3.5 Sonnet & 175B & 200k & 8k & \centering \redcross & \centering \redcross & 04/2024 \\
        Gemini 2.5 Pro & n/a & 1M & 65k & \centering \greentick & \centering \redcross & 01/2025 \\
        Gemini 2.5 Flash & n/a & 1M & 65k & \centering \greentick & \centering \redcross & 01/2025 \\
        Llama 4 Maverick & 400B & 1M & n/a & \centering \redcross & \centering \greentick & 08/2024 \\
        Qwen 3 & 235B & 128k & 64k & \centering \greentick & \centering \greentick & 06/2024 \\
        Deepseek V3 & 671B & 64k & 8k & \centering \redcross & \centering \greentick & 07/2024 \\
        DeepSeek R1 & 671B & 128k & 8k & \centering \greentick & \centering \greentick & 07/2024 \\
        \bottomrule
    \end{tabularx}
    \caption{Studied LLMs.}
    \label{tab:llms}
\end{table}

\descr{Dataset $(D_S)$.} To evaluate the capabilities of LLMs in reproducing vulnerabilities using \cvegenie{}, we construct a small-scale, diverse dataset $(D_S)$ consisting of $60$ CVEs. These CVEs were published after the latest knowledge cutoff date of the LLMs under study (January 2025), as listed in Table~\ref{tab:llms}. To ensure that $D_S$ is both representative and diverse, we select 40 CVEs from the top 25 most dangerous CWE~\cite{mitre--top25_nodate} categories, and the remaining 20 CVEs are randomly sampled from the other CWE types. The resulting dataset, $D_S$, comprises 60 CVEs drawn from 44 different CWE categories, spanning 18 programming languages, and involving 56 distinct software projects.

\descr{Methodology.} \cvegenie{} is composed of four modules. The \cveprocessor's LLM-based component, the \knowledge, only performs a trivial task of summarization, so we assign a lightweight LLM (\texttt{o4-mini}) to it. In contrast, the remaining modules, i.e., \project, \exploit, and \verifier, require strong security vulnerability reasoning and SWE capabilities. Thus, for each of these modules, we systematically select the best performing LLM for ``developer'' and ``critic'' tasks using the following 5-step procedure:

\descrit{(1) Candidate Selection:} For each module, we identify subsets of LLMs from Table~\ref{tab:llms} as potential developers and critics, based on prior evidence of their capabilities, as well as time and cost constraints.

\descrit{(2) Developer Evaluation:} Each candidate developer LLM is run for the given module, and its outputs are manually scored by the authors. The highest-scoring LLM is chosen as the developer.

\descrit{(3) Critic Prompt Design:} We analyze common mistakes made by developers and use these insights to design a module-specific critic prompt, enabling the critic to effectively detect those mistakes.

\descrit{(4) Critic Evaluation:} Each candidate critic LLM is evaluated on true positive rate (TPR) and true negative rate (TNR). When TPRs are equal, we prefer the model with the higher TNR, and vice versa. Then then critic with the best trade-off is selected.

\descrit{(5) Feedback Loop Assessment:} We rerun the chosen best developer–critic pair for the given module, and evaluate the correctness of the critic's feedback and the developer's capability to correctly address the critic's feedback.

\subsubsection{Builder}
\label{subsubsec:llm-project}

In this section, we systematically select the most optimal configuration of LLMs and prompts for the \project.

\begin{figure}[t]
    \centering
    \begin{subfigure}{0.48\linewidth}
        \includegraphics[width=\linewidth]{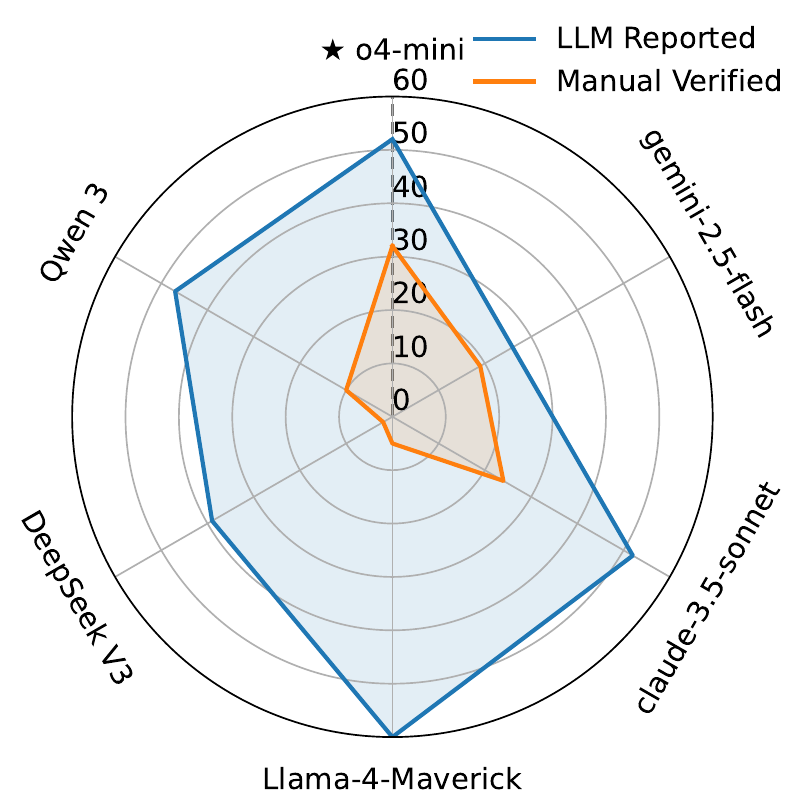}
        \caption{Developer Agent}
        \label{subfig:setup-dev}
    \end{subfigure}
    \hfill
    \begin{subfigure}{0.48\linewidth}
        \includegraphics[width=\linewidth]{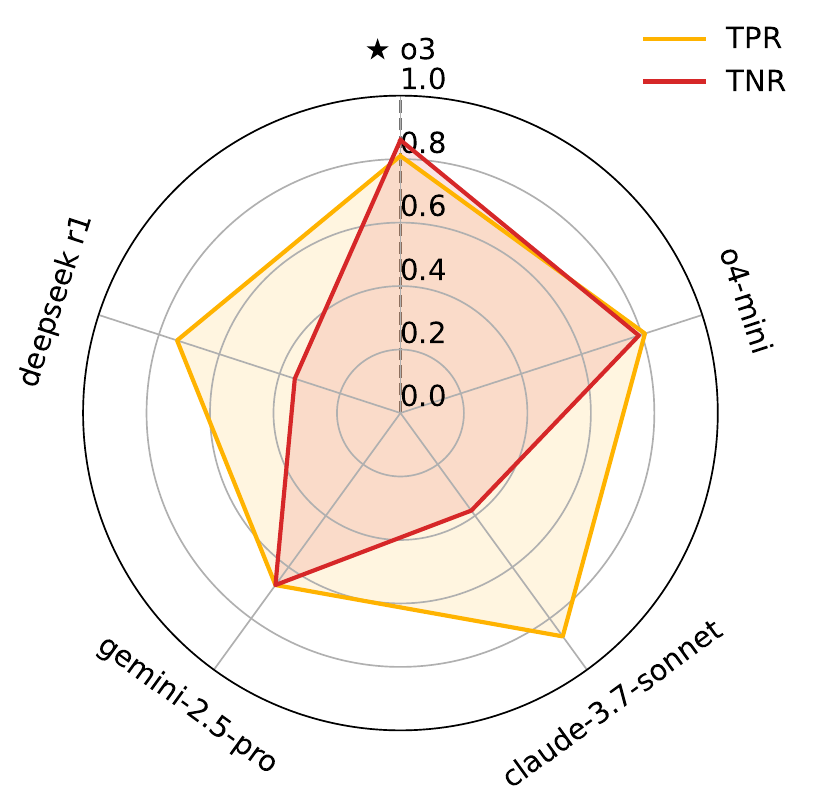}
        \caption{Critic Agent}
        \label{subfig:setup-critic}
    \end{subfigure}
    \caption{Builder Optimal LLM Evaluation}
    \label{fig:llm-project}
\end{figure}

\descrit{Candidate Selection.} Setting up a project from scratch is the most time-consuming and complex part of \cvegenie, as security advisories or CVEs do not provide instructions for setting up the vulnerable project. To address this, we use two \emph{developer} agents: one for exploring the project and another for setting it up (see Section~\ref{subsec:project_builder}). This process requires extensive tool calls and command executions, making it both slow and costly with reasoning-heavy models like \texttt{o3} or \texttt{claude-3.7-sonnet}, which can cost \$10–20 per CVE. Therefore, we select fast, cost-effective LLMs with sufficient context capacity, \texttt{o4-mini}, \texttt{claude-3.5-sonnet}, \texttt{gemini-2.5-flash}, \texttt{llama-4-maverick}, \texttt{deepseek-v3}, and \texttt{qwen3}, as \emph{developer} agent candidates. For \emph{critic} agents, we opt for stronger reasoning models such as \texttt{o3}, \texttt{o4-mini}, \texttt{claude-3.7-sonnet}, \texttt{gemini-2.5-pro}, and \texttt{deepseek-r1}, as the critic's task only requires one call, and therefore, they incur minimal overhead.

\descrit{Developer Evaluation.} We ran \cvegenie{} up to the \setupdev{} stage \cirnum{4} on CVEs from $D_S$ and manually evaluated each developer LLM using the following procedure. \ul{(1) For each CVE, we first identified the official installation documentation for the software’s vulnerable version and, based on it, established both the expected final build state and the sequence of commands required for installation.} \ul{(2) We then reviewed the logs of runs where the LLM reported successful setup, comparing the executed commands against the documented installation process to assess alignment and completeness.} \ul{(3) Finally, we validated the resulting environment: for libraries or binaries, we confirmed the vulnerable version by invoking the corresponding \texttt{--version} (or equivalent) command (e.g., for CVE-2025-1215 we verified that \texttt{vim v9.1.1096} was installed), while for server-based software we performed a health check to ensure the service was running and accessible at the expected port.}

As shown in Figure~\ref{subfig:setup-dev}, \texttt{o4-mini} emerged as the best developer LLM for the \project, successfully setting up 32 out of 60 CVEs projects, while making an average of 20 tool calls per CVE. In contrast, \texttt{claude-3.5-sonnet} mostly gave up on the given setup task, and models like \texttt{gemini-2.5-flash} and open-source LLMs were unreliable: \texttt{Qwen 3} often described setup steps without executing them using tool calls, while \texttt{gemini-2.5-flash} issued tool calls with frequent syntax errors, eventually leading to the failed setup of the given project.

\descrit{Critic Prompt Design.} From analyzing \emph{developer} LLM setups, we observed three recurring mistakes: (1) when setup fails, the LLM builds a simplified mock-up substitute project instead; (2) for \texttt{pip}/\texttt{npm} packages, it installs the latest version instead of the specified vulnerable one; and (3) when a server is required, it assumes commands like \texttt{npm run dev} succeed without verifying server health. We designed the critic prompt to detect these errors while reviewing \setupdev{} logs.

\descrit{Critic Evaluation.} As shown in Figure \ref{subfig:setup-critic}, the \texttt{o3} model achieves the highest TPR and TNR when evaluating \setupdev's execution logs. Most of its false negatives involved setups requiring external hardware, which \exploitdev{} cannot exploit anyway. Some valid setups were also mistakenly rejected due to limited post-setup checks. Given its strong balance of recall and precision, we choose \texttt{o3} as the critic model for the \project.

\descrit{Feedback Assessment.} We evaluated \cvegenie{} up to \setupdev{} stage \cirnum{4}, using the best-performing developer (\texttt{o4-mini}) and critic (\texttt{o3}) models. Minor issues flagged by the \emph{critic} (e.g., limited verification) were typically resolved by the \emph{developer} in one iteration. However, fundamental issues (e.g., mock-up version of the project) persisted even after five iterations. Therefore, we limit feedback to a single iteration to efficiently address minor issues without incurring excessive overhead.

\subsubsection{Exploiter}
\label{subsubsec:llm-exploit}

In this section, we systematically select the most optimal configuration of LLMs and prompts for the \exploit.

\begin{figure}[t]
    \centering
    \begin{subfigure}{0.48\linewidth}
        \includegraphics[width=\linewidth]{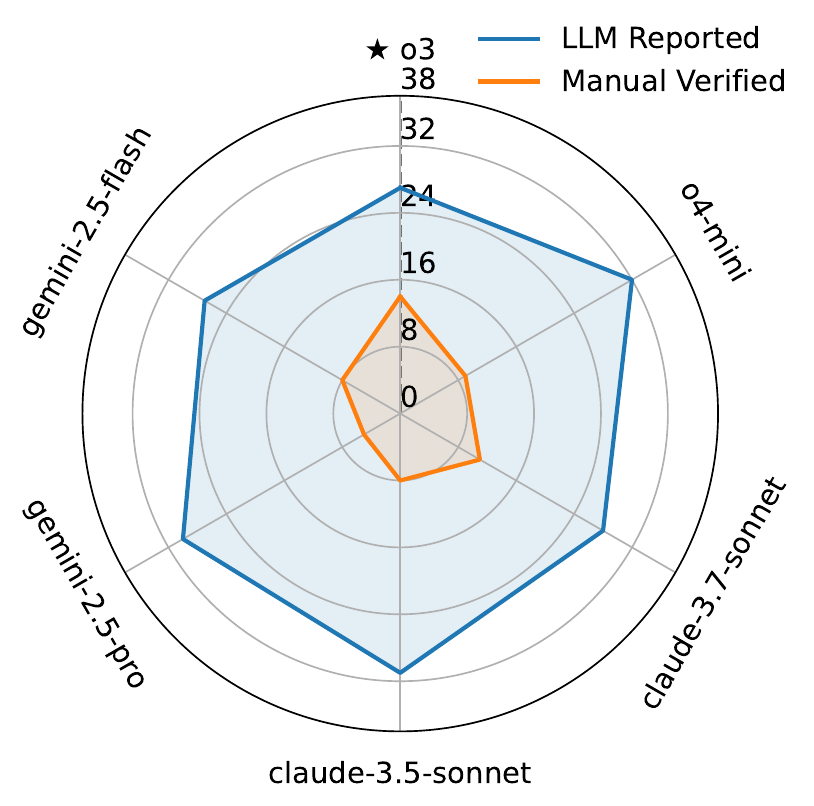}
        \caption{Developer Agent}
        \label{subfig:exploit-dev}
    \end{subfigure}
    \hfill
    \begin{subfigure}{0.48\linewidth}
        \includegraphics[width=\linewidth]{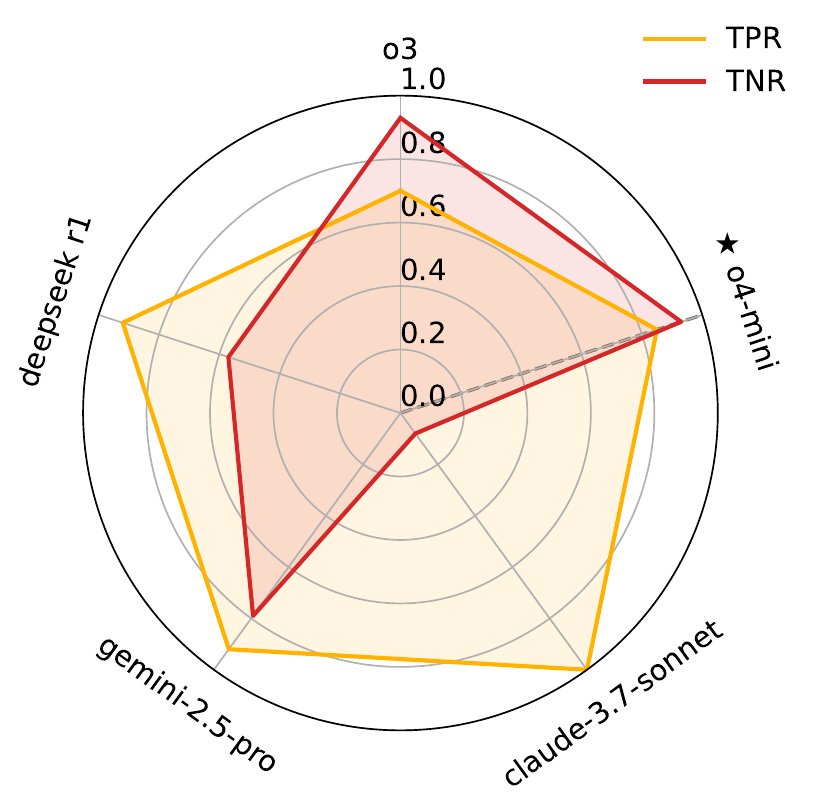}
        \caption{Critic Agent}
        \label{subfig:exploit-critic}
    \end{subfigure}
    \caption{Exploiter Optimal LLM Evaluation}
    \label{fig:llm-exploit}
\end{figure}

\descrit{Candidate Selection.} Generating a working exploit from a given CVE in a pre-configured vulnerable environment is a crucial task for reproducing CVEs. However, the \exploit{} requires less code base exploration than the \project, as advisories often provide PoCs, exploitation steps, or vulnerability details. Thus, we include expensive reasoning models like \texttt{o3}, \texttt{claude-3.7-sonnet}, and \texttt{gemini-2.5-pro} in the candidates list of developers. To mitigate instability seen in open-source models (Section~\ref{subsubsec:llm-project}), we employ only closed-source LLMs as developers, while using the same critic LLMs as the \project.

\descrit{Developer Evaluation.} For the 32 successful setups from Section~\ref{subsubsec:llm-project}, we ran \cvegenie{} up to the \exploitdev{} stage \cirnum{6} and manually evaluated each developer LLM as follows: \ul{(1) we checked whether the LLM executed a PoC or command sequence that demonstrated the exploit, (2) we verified that the vulnerability was clearly triggered (e.g., uncontrolled recursion in CWE-674 resulting in a \texttt{RecursionError}), (3) we assessed fidelity to the CVE description, (4) we ensured no mock-up environments or fake exploits were used, and (5) we confirmed the exploit script followed the expected format} (see Section~\ref{subsubsec:exp-dev}).
Among all models, \texttt{o3} consistently performed best, successfully reproducing the highest number of exploits (13 out of 27) and demonstrating them end-to-end within the actual vulnerable environment. In contrast, \texttt{o4-mini} often produced incorrect or mock-up exploits and struggled with runtime errors. While \texttt{claude} and \texttt{gemini} models showed strong capabilities in demonstrating exploits, they lacked versatility across different vulnerability types. Hence, we select \texttt{o3} as the \exploitdev.

\descrit{Critic Prompt Design.} We examined the failed exploit attempts and categorized the recurring error patterns, which directly mirrored our manual verification criteria (e.g., lack of exploit demonstration, unclear vulnerability trigger, deviation from CVE description, reliance on dummy/fake exploits, or incorrect script formatting). We then incorporated these categories into the \emph{critic} agent’s prompt, enabling it to automatically detect these issues for \exploitdev{}.

\descrit{Critic Evaluation.} As shown in Figure~\ref{subfig:exploit-critic}, \texttt{o4-mini} achieved the best balance of recall and specificity. In contrast, \texttt{o3} was overly strict, rejecting correct exploits and focusing too much on code formatting, while \texttt{gemini-2.5-pro} frequently overlooked incomplete exploit verification. \texttt{Claude-3.7-sonnet} was too lenient, accepting nearly all exploits. Based on this evaluation, we selected \texttt{o4-mini} as the most effective critic model for the \exploitcritic.

\descrit{Feedback Assessment.} We evaluated feedback effectiveness by running \cvegenie{} on 32 correct setups with the top developer (\texttt{o3}) and critic (\texttt{o4-mini}). This yielded valid exploits for 16 CVEs, of which 3 were fixed based on the critic’s feedback. As in Section~\ref{subsubsec:llm-project}, the developer mainly resolved minor issues (e.g., incomplete verification or missing log triggers), while more complex flaws (e.g., incorrect exploit verification) persisted. To balance usefulness and cost, we therefore restrict feedback to a single iteration.

\subsubsection{Verifier}
\label{subsubsec:llm-verify}

In this section, we systematically select the most optimal configuration of LLMs and prompts for the \verifier.

\begin{figure}[t]
    \centering
    \begin{subfigure}{0.48\linewidth}
        \includegraphics[width=\linewidth]{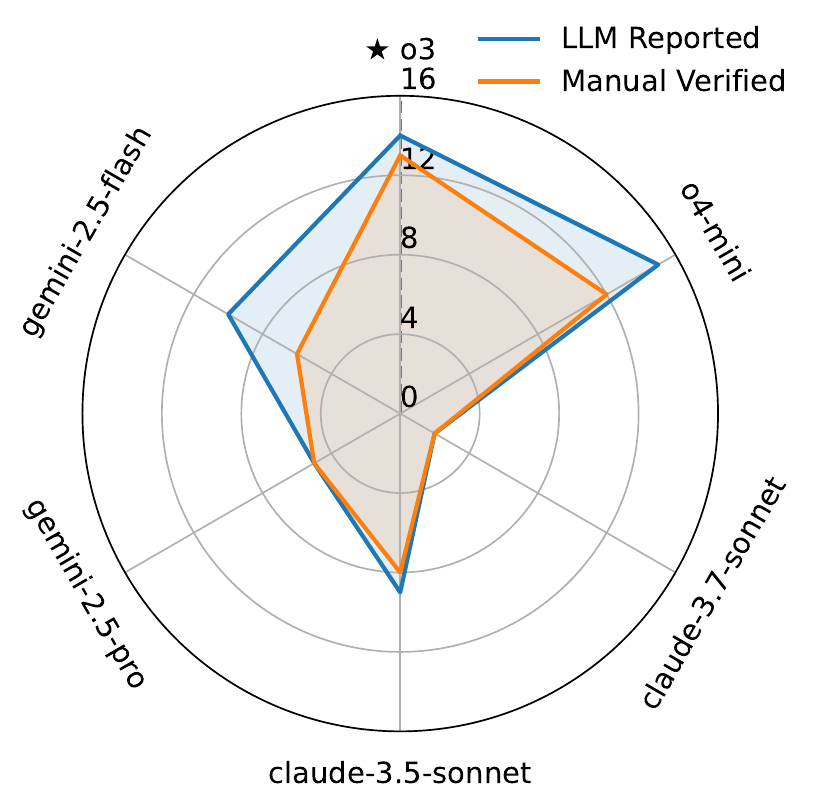}
        \caption{Developer Agent}
        \label{subfig:verifier-dev}
    \end{subfigure}
    \hfill
    \begin{subfigure}{0.48\linewidth}
        \includegraphics[width=\linewidth]{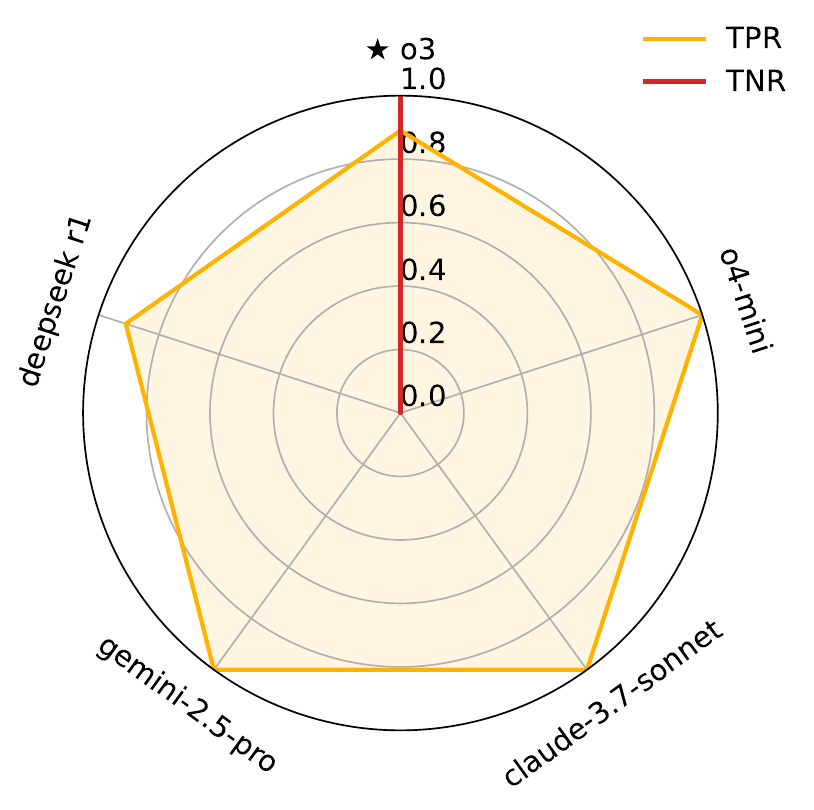}
        \caption{Critic Agent}
        \label{subfig:verifier-critic}
    \end{subfigure}
    \caption{Verifier Optimal LLM Evaluation}
    \label{fig:llm-verifier}
\end{figure}

\descrit{Candidate Selection.} Like exploit generation, creating a verifier also demands deep understanding of security vulnerabilities to ensure the exploit is successfully triggered. Therefore, we use the same developer and critic candidates as in the \exploit.

\descrit{Developer Evaluation.} For the 16 valid exploits from Section~\ref{subsubsec:llm-exploit}, we ran \cvegenie{} up to the \flagcheck{} stage \cirnum{9} and manually evaluated each developer LLM on three criteria: \ul{(1) whether the verifier script followed the required format} (see Section~\ref{subsubsec:ctf-ver-dev}), \ul{(2) whether it executed the provided exploit script without modification, and (3) whether its verification logic was precise and reliable} (see Figure~\ref{fig:verifier-example} for a detailed example).

As shown in Figure~\ref{subfig:verifier-dev}, the \texttt{o3} model generated 13 valid verifiers out of 14, outperforming all others. While \texttt{o4-mini} achieved comparable numbers, its verifiers more often relied on weak logic. Other models struggled with functional correctness and error recovery. We therefore selected \texttt{o3} for the \verifierdev{} due to its consistent one-shot success in producing correct verifiers.

\descrit{Critic Prompt Design.} We analyzed failures in verifier generation and found that the recurring error types matched our verification criteria (e.g., incomplete checks, incorrect validation logic, or misuse of the target environment). We used these categories to design the critic agent’s prompt, enabling it to automatically flag such errors when evaluating verifiers produced by the \verifierdev.

\descrit{Critic Evaluation.} As shown in Figure~\ref{subfig:verifier-critic}, only the \texttt{o3} model successfully identified critical but subtle issues in the verifier scripts, such as weak or flawed verification logic (Figure~\ref{fig:verifier-example}). Other models provided weak or no critique, making \texttt{o3} the most %

\descrit{Feedback Assessment.} We evaluated \cvegenie{} on 16 correct exploits from the \exploit, successfully generating accurate verifiers for 15 of them, of which 4 verifiers were improved through the critic's feedback. In our experiment, the maximum number of feedback iterations required to resolve either functional or critique-based issues was 4; therefore, we allow up to 5 retries for both the \flagcheck{} and the \verifiercritic.

\subsubsection{Time and Cost Constraints}
\label{subsubsection:time-cost}

In our evaluation of \cvegenie, we found that successfully reproduced CVEs typically required about \$2 and 18 minutes on average, with the most resource-intensive case costing \$6 and taking 45 minutes. Failed reproductions, on the other hand, averaged up to \$4 and 35 minutes. To balance flexibility with cost control, we set a per-CVE budget cap of \$5 and a maximum runtime of 45 minutes for all experiments.

\section{Experimental Methodology}
\label{sec:exp}
In this section, we conduct a four-part evaluation of \cvegenie{}: (1) a baseline study assessing consistency and efficiency (Section~\ref{subsec:sat-study}); (2) an ablation study to assess the impact of key components of \cvegenie{} on its scalability and effectiveness (Section~\ref{subsec:design-ablation}); (3) robustness testing under incomplete CVE information (Section~\ref{subsec:robust}); and (4) large-scale evaluation on 841 CVEs (Section~\ref{subsec:large-scale}). These evaluations are guided by the following research questions:

\begin{compactitem}
    \item \textbf{RQ1:} Does the performance of \cvegenie{} vary over multiple runs of CVE reproduction?
    
    \item \textbf{RQ2:} Is the complex architecture of \cvegenie{} necessary, or can standalone LLMs achieve comparable results?
    
    \item \textbf{RQ3:} Can \cvegenie{} effectively reproduce CVEs with limited information, and which CVE report components are most crucial for reproduction?
    
    \item \textbf{RQ4:} Can \cvegenie{} reproduce a broad range of CVEs across diverse CWEs, languages, and projects?
\end{compactitem}

\subsection{Baseline Assessment}
\label{subsec:sat-study}

In this section, we evaluate the baseline performance of \cvegenie{} and examine how its reproduction success evolves across multiple runs in the presence of LLM non-determinism.

\descr{Dataset.} For this study we use the same $D_S$ dataset (Section~\ref{subsec:llms-cve-genie}).

\descr{Methodology.} Due to the non-deterministic nature of LLMs, a single run may fail due to incorrect reasoning, while later attempts might succeed~\cite{manvi_adaptive--inf_2024}. To account for this, we run \cvegenie{} multiple times on $D_S$, using its most optimal configuration (Section~\ref{subsec:llms-cve-genie}). After each iteration, we store successfully reproduced CVEs and rerun on the remaining ones. This iterative approach helps maximize CVE reproduction and reveals \cvegenie's convergence.

\begin{figure}[t]
    \centering
    \includegraphics[width=\linewidth]{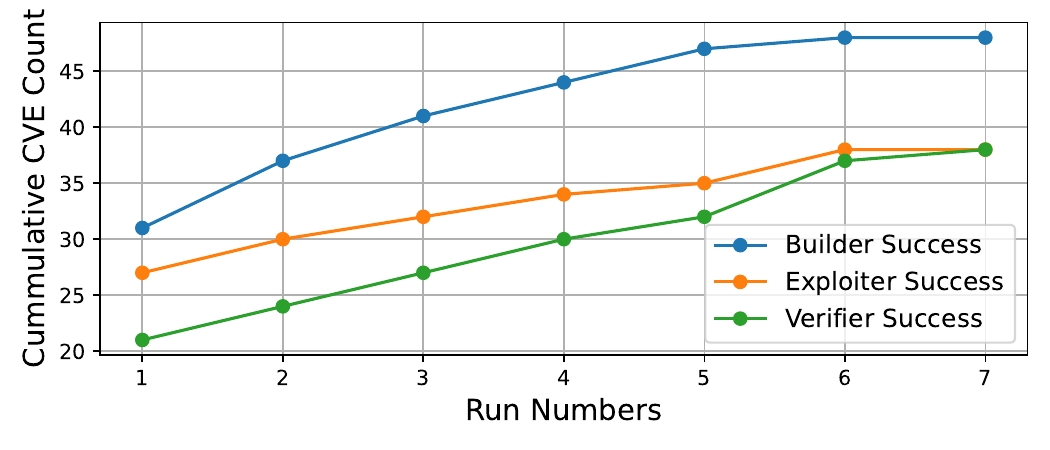}
    \caption{\cvegenie{} performance over five runs on $D_S$.}
    \label{fig:sat-exp}
\end{figure}

\descr{Results.} We perform manual analysis of randomly selected 25 CVE reproduction runs, and observe that the variability in \cvegenie's performance primarily stems from the \project ~phase, which is more open-ended than the targeted exploit generation guided by CVE context. Project setup involves repository exploration (\predev) and command execution (\setupdev), where environment-specific behaviors often cause failures. For example, one run failed because the \setupdev{} modified the \texttt{PATH} variable, while a clean retry succeeded, highlighting the value of reattempting in a fresh environment. Since the \verifier{} depends on the \exploit{}, which in turn relies on the \project{}, this initial project setup is a critical bottleneck. As a previously failed build, if later successful, can enable downstream modules to reproduce a new CVE (as illustrated in Figure~\ref{fig:sat-exp}). Over multiple runs, the number of successful project, exploit, and verifier builds showed consistent gains with convergence at run seven, indicating that most recoverable failures can be resolved within a few retries.

\begin{researchbox}
\textbf{RQ1:} Due to the non-determinism of LLMs, \cvegenie{} exhibits variability across runs, primarily during project setup and environment configuration. Once project setup succeeds, downstream exploit generation and verification are more stable and converge in a few retries.
\end{researchbox}

\subsection{Importance of \cvegenie's Architecture}
\label{subsec:design-ablation}

Having evaluated the selection of LLMs and prompts for \cvegenie{} (Section~\ref{subsec:llms-cve-genie}), we now assess \cvegenie{}’s architecture via a systematic ablation study that isolates the impact of \cvegenie's individual components on end-to-end CVE reproduction.

\descr{Dataset.} 
We perform this ablation study using 15 CVEs from $D_S$, each with complete information (i.e., CVE description, PoC, patch commit, and security advisory) and reproduced consistently in the first three iterations of baseline assessment (Section~\ref{subsec:sat-study}), and call this dataset $D_R$. By keeping the CVE context complete, we ensure that any performance differences arise solely from modifications%

\descr{Methodology.} As \dataprocessor, \setupdev, \exploitdev, and \verifierdev{} are the four indispensable components of \cvegenie{} for end-to-end CVE reproduction, for this evaluation of \cvegenie's design we use the following five ablation settings:

\begin{compactenum}
    \item \textit{No \knowledge:} Bypass the Knowledge Builder and feed raw data directly to all agents.
    \item \textit{No \predev:} Eliminate the \predev, allowing the \setupdev{} to independently plan and execute the entire environment setup from scratch.
    \item \textit{No Feedback Loops:} Enforce single-shot execution without iterative refinement, by removing all feedback loops.
    \item \textit{No Critics:} Remove the \verifiercritic{} and treat a CVE as reproduced if the final scripts pass the \flagcheck.
    \item \textit{Single Monolithic Agent:} Combine all modular agents into a single agent with access to the same tools, thereby evaluating the performance of a standalone LLM without \cvegenie{}'s agentic design and structured guidance (see Appendix~\ref{sec:single-agent}). %
\end{compactenum}

\descr{Results.}
In our ablation study (Table~\ref{tab:design-ablation}), removing the \knowledge{} component reduces reproduction success to 9/15, as agents struggle to interpret unstructured advisory data and hit context window limits more often, which results in a 30\% increase in exploit failures. Eliminating the \predev{} has a comparatively mild effect (13/15), but places additional burden on the \setupdev{} as it has to do both codebase exploration and project setup at the same time, leading to 27\% more tool-call limit violations. We observe this behavior particularly in complex repositories such as \texttt{CVE-2025-32389} in \texttt{Nameless}, a website software for Minecraft servers, with not very clear setup guides.
Feedback loops emerged as the most critical design element, and without them the reproduction success dropped sharply to 5/15 (–67\%). Similarly, removing critic agents reduced success to 8/15 while increasing false reproductions by 47\%, highlighting their role in reliable end-to-end CVE reproduction. Finally, collapsing all components into a single monolithic agent resulted in zero successful reproductions, even when using advanced standalone LLMs (\texttt{o3}), underscoring the necessity of \cvegenie{}’s modular, feedback-driven, and critic-guided %

\begin{researchbox}
\textbf{RQ2:} Standalone LLMs are currently unable to reproduce CVEs end-to-end. \cvegenie{}’s modular, feedback-driven, and critic-guided architecture design is essential for reliable CVE reproduction, and removing any of its components significantly degrades performance (Table~\ref{tab:design-ablation}).
\end{researchbox}

\begin{table}[h]
    \small
    \centering
    \rowcolors{2}{gray!10}{white}
    \begin{tabularx}{\linewidth}{>{\centering\arraybackslash}X|>{\centering\arraybackslash}X>{\centering\arraybackslash}X>{\centering\arraybackslash}X>{\centering\arraybackslash}X|>{\centering\arraybackslash}X}
        \toprule
        \textbf{Ideal Case} & \textbf{Knw Build} & \textbf{Pre- Reqs} & \textbf{Feed-back} & \textbf{Critic Agents} & \textbf{Single Agent} \\
        \midrule
        15 {\scriptsize / 15} & 9 {\scriptsize / 15} & 13 {\scriptsize / 15} & 5 {\scriptsize / 15} & 8 {\scriptsize / 15} & 0 {\scriptsize / 15} \\
        \midrule
        - & $\uparrow$ 30\% Exploit Failure & $\uparrow$ 27\% Max Tool & $\downarrow$ 67\% Reprod Rate & $\uparrow$ 47\% False Reprod & - \\
        \bottomrule
    \end{tabularx}
    \caption{Ablation study of \cvegenie{}'s design.}
    \label{tab:design-ablation}
\end{table}

\subsection{Robustness to Loss of CVE Context}
\label{subsec:robust}

To evaluate how \cvegenie{} performs when critical components of a CVE report are missing, we study its robustness under systematically reduced CVE context. We first present a detailed \emph{qualitative} case study of a single CVE to illustrate how \cvegenie{} adapts its reasoning under different context losses, and then conduct a \emph{quantitative} ablation study to quantify the impact at scale.

\subsubsection{Case Study for CVE-2024-4340.} 
\label{subsubsec:case-study-cve-context}

We begin by performing a detailed case study for CVE-2024-4340 reproduction under three levels of CVE context losses, and compare their reproductions against the ideal setting shown in Figure~\ref{fig:architecture}.

\descrit{1) No Security Advisory / PoC.} In this setting, all security advisories are removed from the CVE context, leaving only the developer patch commit\footnote{CVE-2024-4340 Developer Patch: \url{https://github.com/andialbrecht/sqlparse/commit/b4a39d9850969b4e1d6940d32094ee0b42a2cf03}} and the CVE description\footnote{CVE-2024-4340 CVE Description: ``Passing a heavily nested list to sqlparse.parse() leads to a Denial of Service due to RecursionError.''}. As a result, \cvegenie{} relies heavily on analyzing the patch commit, exploring modified files, and inferring the PoC from the regression test added by the developer in \verb|tests/test_regressions.py|. This test closely resembles the PoC described in the existing advisory\footnote{CVE-2024-4340 GHSA: \url{https://github.com/advisories/GHSA-2m57-hf25-phgg}}, and the generated exploit payload (Figure~\ref{subfig:adv-patch-payload}) is comparable to the ideal run shown in Figure~\ref{subfig:exploit-script}. Due to the absence of explicit PoC information, this run required additional exploration, resulting in a longer execution time (9 minutes) and higher cost (\$0.67) compared to the ideal scenario.

\descrit{2) No Patch.} Here, the developer patch commit is removed while security advisories remain available. \cvegenie{} primarily leverages the PoC provided in the advisory, which includes clear PoC exploit code, execution traces, vulnerable code, and vulnerability root cause. Consequently, the reproduction closely matches the ideal run in terms of generated PoC exploit (Figure~\ref{subfig:adv-patch-payload}), verification scripts, execution time, and cost.

\descrit{3) Only CVE Description.} In the most constrained setting, \cvegenie{} has access only to the CVE description. Based on this limited information, it starts by identifying the implementation of \texttt{sqlparse.parse} and hypothesizes on how a deeply nested inputs can trigger a recursive execution path leading to a \verb|RecursionError|. During PoC exploit generation and verification, the first run failed to produce a payload that successfully triggered the \verb|RecursionError|. However, a second run from scratch succeeded in generating a valid payload (shown in Figure~\ref{subfig:only-desc-payload}). This setting proved to be the most challenging, requiring 20 minutes of execution time and \$4.50 LLM cost over two runs.

\begin{figure}[t]
\centering

\begin{subfigure}{\linewidth}
\centering
\lstset{
    language=Python,
    basicstyle=\footnotesize\ttfamily,
    backgroundcolor=\color{gray!10},
    numbers=left,
    numberstyle=\tiny\color{gray}\ttfamily,
    stepnumber=1,
    numbersep=5pt,
    showspaces=false,
    showstringspaces=false,
    showtabs=false,
    frame=none,
    framesep=0pt,
    tabsize=4,
    breaklines=true,
    postbreak=\mbox{\textcolor{gray}{$\hookrightarrow$}\space},
    breakatwhitespace=false,
    columns=fullflexible,
    keepspaces=true,
    xleftmargin=0pt,
    xrightmargin=0pt,
    aboveskip=0pt,
    belowskip=0pt,
    lineskip=0pt,
    keywordstyle=\color{green!50!black}\bfseries,
    commentstyle=\color{red!70!black}\itshape,
    stringstyle=\color{red!70!black},
}
\setlength{\fboxsep}{0pt}
\begin{lstlisting}[escapechar=@]
@\colorbox{cyan!17}{\strut\hspace{-\fboxsep}\makebox[\linewidth][l]{payload~=~\textcolor{red!70!black}{\texttt{'['}}~*~depth~+~\textcolor{red!70!black}{\texttt{']'}}~*~depth}}@
sqlparse.parse(payload)
\end{lstlisting}
\caption{Security advisory/PoC or patch commit is not available.}
\label{subfig:adv-patch-payload}
\end{subfigure}

\hfill

\begin{subfigure}{\linewidth}
\centering
\lstset{
    language=Python,
    basicstyle=\footnotesize\ttfamily,
    backgroundcolor=\color{gray!10},
    numbers=left,
    numberstyle=\tiny\color{gray}\ttfamily,
    stepnumber=1,
    numbersep=5pt,
    showspaces=false,
    showstringspaces=false,
    showtabs=false,
    frame=none,
    framesep=0pt,
    tabsize=4,
    breaklines=true,
    postbreak=\mbox{\textcolor{gray}{$\hookrightarrow$}\space},
    breakatwhitespace=false,
    columns=fullflexible,
    keepspaces=true,
    xleftmargin=0pt,
    xrightmargin=0pt,
    aboveskip=0pt,
    belowskip=0pt,
    lineskip=0pt,
    keywordstyle=\color{green!50!black}\bfseries,
    commentstyle=\color{red!70!black}\itshape,
    stringstyle=\color{red!70!black},
}
\setlength{\fboxsep}{0pt}
\begin{lstlisting}[escapechar=@]
@\colorbox{cyan!17}{\strut\hspace{-\fboxsep}\makebox[\linewidth][l]{payload~=~\textcolor{red!70!black}{\texttt{'select'}}~+~\textcolor{red!70!black}{\texttt{'('}}~*~depth~+~\textcolor{red!70!black}{\texttt{'1'}}~+~\textcolor{red!70!black}{\texttt{')'}}~*~depth}}@
sqlparse.parse(payload)
\end{lstlisting}
\caption{Only CVE description is available.}
\label{subfig:only-desc-payload}
\end{subfigure}

\caption{PoC exploit payload generated by \cvegenie{} for CVE-2024-4340 under varying levels of incomplete CVE context compared to the payload in ideal scenario in Figure~\ref{subfig:exploit-script}.}
\label{fig:payloads}
\end{figure}

\subsubsection{Evaluation at Scale.}
\label{subsubsec:scale-eval-cve-context}

In this section, we perform an ablation study to systematically quantify the robustness of \cvegenie{} to incomplete CVE context at scale.

\descrit{Dataset ($D_R$)}. We perform this ablation with $D_R$ dataset (Section~\ref{subsec:design-ablation}).
We intentionally select these ``easier'' CVEs as a controlled probe, since \cvegenie{} can consistently reproduce them under full context.
Hence, any performance degradation after removing specific information indicates that the missing context is genuinely critical. 
To simulate incomplete CVE context, we construct three variants of $D_R$ by systematically removing: (a) security advisory/PoC, (b) patch commit, and (c) everything except CVE description.

\descrit{Methodology.} In this study, for each variant of $D_R$, we run \cvegenie{} three times per CVE, iteratively.

\descrit{Results.} 
Removing the developer patch commit leads to a notable drop in reproduction success 10/15 (67\%), even when a PoC is available. Patch commits provide essential localization signals by identifying relevant files and code paths. Without patch commits, the model must rely on broad codebase exploration, often resulting in inefficient search and budget exhaustion. We observed this behavior in CVE-2025-31481 reproduction, an authorization flaw in the PHP-based \texttt{API Platform Core}, where the absence of a patch caused \cvegenie{} to explore unrelated components and fail to converge.
Excluding security advisories reduces the success rate to 11/15 (73\%). In these cases, failures are primarily attributable to environment setup and verification difficulties rather than exploit generation. Advisories often provide crucial contextual information, such as dependency versions, execution constraints, and validation guidance, that streamline reproduction, particularly when PoCs are not provided as executable code but rather as high-level textual descriptions.
Despite these challenges, \cvegenie{} maintains baseline robustness under minimal context. When only the CVE description is provided, it successfully reproduces 9/15 cases (60\%), albeit with increased cost and runtime. This demonstrates that while comprehensive CVE reports significantly improve efficiency and reliability, meaningful reproduction remains feasible even %

\begin{researchbox}
\textbf{RQ3:} \cvegenie{} can reproduce CVEs under limited context. \emph{Code-based PoCs} provide the strongest guidance, but in their absence, \emph{patch commits} help localize vulnerabilities. Reproduction remains possible even with only \emph{CVE descriptions}, but with lower success rates, higher cost, and increased time.
\end{researchbox}

\begin{figure}[t]
    \centering
    \includegraphics[width=\linewidth]{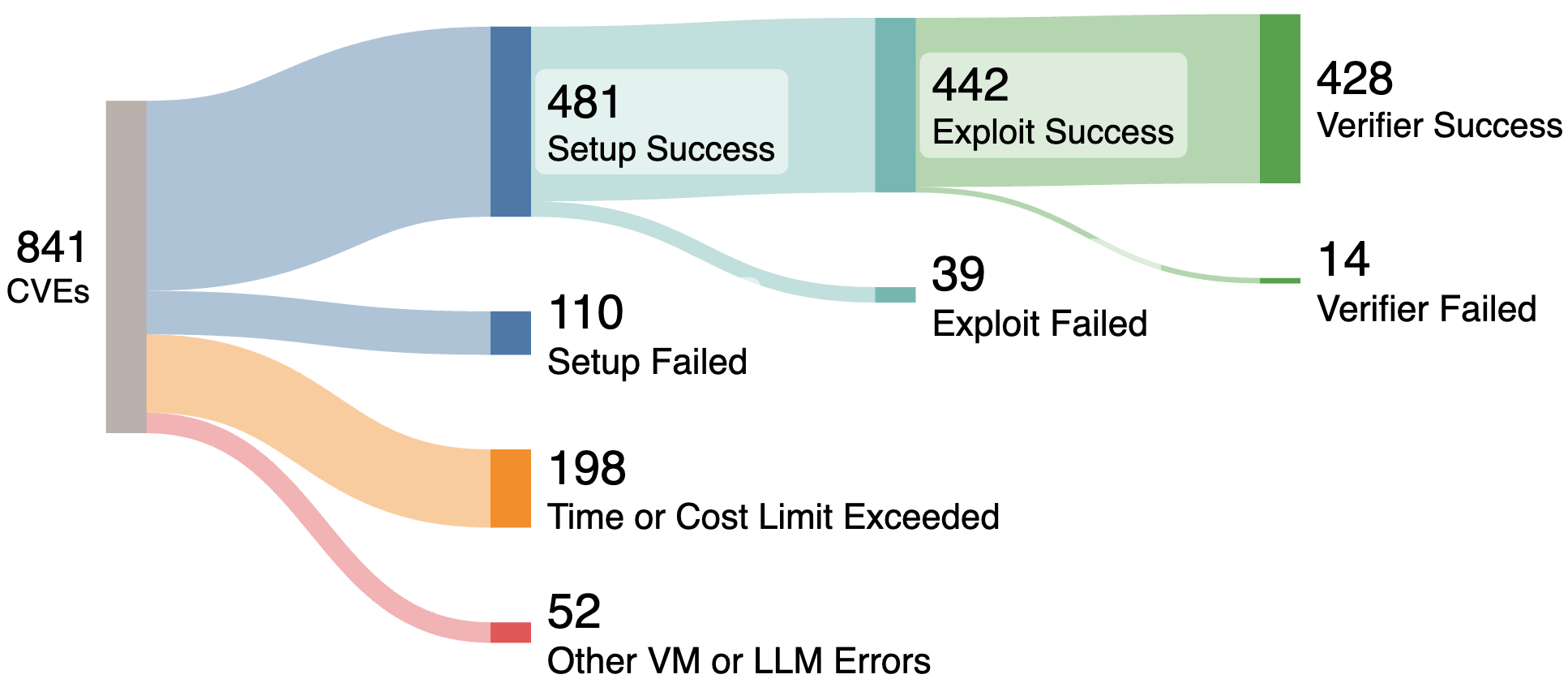}
    \caption{Breakdown of \cvegenie's run on $D_L$}
    \label{fig:large-scale}
\end{figure}

\begin{table}[b]
    \centering
    \footnotesize
    \rowcolors{2}{gray!10}{white}
    \begin{tabularx}{\linewidth}{
        lrr
    }
        \toprule
        \textbf{Project Type} &
        \textbf{CVEs Reprod.} &
        \textbf{Projects Reprod.} \\
        \midrule
        AI/ML Platform              & \textbf{22/38 (57.9\%)}   & \textbf{14/16 (87.5\%)}   \\
        Web Application/Backend     & \textbf{171/383 (44.6\%)} & \textbf{86/156 (55.1\%)}  \\
        Library/Framework           & \textbf{163/224 (72.8\%)} & \textbf{115/147 (78.2\%)} \\
        Cloud/DevOps Application  & 13/33 (39.4\%)   & 13/25 (52.0\%)   \\
        Operating System/Runtime    & 15/31 (48.4\%)   & 5/15 (33.3\%)    \\
        Desktop Application         & 6/34 (17.6\%)    & 2/13 (15.4\%)    \\
        CLI Tool/Utility            & \textbf{18/37 (48.6\%)}   & \textbf{15/28 (53.6\%)}   \\
        Blockchain/Crypto/FinTech   & 1/13 (7.7\%)     & 1/7 (14.3\%)     \\
        Mobile Application/SDK      & 2/7 (28.6\%)     & 2/7 (28.6\%)     \\
        Embedded/Networking         & \textbf{11/27 (40.7\%)}   & \textbf{10/18 (55.6\%)}   \\
        Security Tool/Server        & \textbf{6/12 (50.0\%)}    & \textbf{4/6 (66.7\%)}     \\
        \bottomrule
    \end{tabularx}
    \caption{CVEs reproduction coverage by project type.}
    \label{tab:repro-summary}
\end{table}

\begin{table}[t]
    \centering
    \footnotesize
    \rowcolors{2}{gray!10}{white}
    \begin{tabularx}{\linewidth}{lrrr}
        \toprule
        \textbf{CWE-ID} &
        \textbf{Overall success} &
        \textbf{PoC available} &
        \textbf{No PoC} \\
        \midrule
        CWE-79   & 65/142 (45.8\%) & 43/94 (45.7\%) & 22/48 (45.8\%) \\
        CWE-200  & 20/43 (46.5\%)  & 15/20 (75.0\%) & 5/23 (21.7\%)  \\
        CWE-22   & 16/37 (43.2\%)  & 11/22 (50.0\%) & 5/15 (33.3\%)  \\
        CWE-284  & 19/36 (52.8\%)  & 16/26 (61.5\%) & 3/10 (30.0\%)  \\
        CWE-400  & 22/36 (61.1\%)  & 13/19 (68.4\%) & 9/17 (52.9\%)  \\
        CWE-20   & 13/29 (44.8\%)  & 6/15 (40.0\%)  & 7/14 (50.0\%)  \\
        CWE-770  & 14/24 (58.3\%)  & 7/9 (77.8\%)   & 7/15 (46.7\%)  \\
        CWE-918  & 7/22 (31.8\%)   & 6/13 (46.2\%)  & 1/9 (11.1\%)   \\
        CWE-89   & 11/21 (52.4\%)  & 8/15 (53.3\%)  & 3/6 (50.0\%)   \\
        CWE-1333 & 15/19 (78.9\%)  & 8/10 (80.0\%)  & 7/9 (77.8\%)   \\
        CWE-94   & 9/19 (47.4\%)   & 2/8 (25.0\%)   & 7/11 (63.6\%)  \\
        CWE-863  & 7/15 (46.7\%)   & 4/6 (66.7\%)   & 3/9 (33.3\%)   \\
        CWE-287  & 8/15 (53.3\%)   & 4/4 (100.0\%)  & 4/11 (36.4\%)  \\
        CWE-285  & 7/15 (46.7\%)   & 5/8 (62.5\%)   & 2/7 (28.6\%)   \\
        CWE-269  & 7/13 (53.8\%)   & 4/8 (50.0\%)   & 3/5 (60.0\%)   \\
        CWE-122  & 8/13 (61.5\%)   & 2/4 (50.0\%)   & 6/9 (66.7\%)   \\
        CWE-502  & 6/13 (46.2\%)   & 3/6 (50.0\%)   & 3/7 (42.9\%)   \\
        CWE-532  & 6/12 (50.0\%)   & 2/2 (100.0\%)  & 4/10 (40.0\%)  \\
        CWE-78   & 4/11 (36.4\%)   & 4/9 (44.4\%)   & 0/2 (0.0\%)    \\
        CWE-352  & 8/10 (80.0\%)   & 5/6 (83.3\%)   & 3/4 (75.0\%)   \\
        CWE-74   & 7/10 (70.0\%)   & 4/4 (100.0\%)  & 3/6 (50.0\%)   \\
        CWE-416  & 3/10 (30.0\%)   & 0/0 (--)       & 3/10 (30.0\%)  \\
        CWE-639  & 2/9 (22.2\%)    & 1/4 (25.0\%)   & 1/5 (20.0\%)   \\
        CWE-116  & 4/8 (50.0\%)    & 1/3 (33.3\%)   & 3/5 (60.0\%)   \\
        CWE-347  & 7/8 (87.5\%)    & 4/4 (100.0\%)  & 3/4 (75.0\%)   \\
        \bottomrule
    \end{tabularx}
    \caption{Top 25 CWEs identified in the large-scale study and their reproduction success rates, overall and stratified by PoC availability.}
    \label{tab:cwe-success}
\end{table}

\begin{table}[t]
    \centering
    \footnotesize
    \rowcolors{2}{gray!10}{white}
    \begin{tabularx}{\linewidth}{l>{\centering\arraybackslash}X>{\centering\arraybackslash}X>{\centering\arraybackslash}X>{\centering\arraybackslash}Xl}
        \toprule
        \textbf{Project} & \textbf{\# Re CVEs} & \textbf{\# T25 CWEs} & \textbf{\# Oth CWEs} & \textbf{LoC} & \textbf{Lang} \\
        \noalign{\vskip 1ex}
        \midrule
        lunary-ai & 35 & 13 & 22 & \textasciitilde70k & TypeScript \\
        WeGIA & 17 & 15 & 2 & \textasciitilde1M & PHP \\
        cpython & 10 & 2 & 8 & \textasciitilde1.8M & Py, C \\
        vite & 6 & 6 & 0 & \textasciitilde107k & TypeScript \\
        yeswiki & 6 & 6 & 0 & \textasciitilde234k & PHP \\
        symfony & 5 & 4 & 1 & \textasciitilde1.6M & PHP \\
        llama\_index & 5 & 4 & 1 & \textasciitilde1.1M & Py \\
        vim & 5 & 3 & 2 & \textasciitilde1.2M & Vim Script, C \\
        directus & 5 & 2 & 3 & \textasciitilde398k & TypeScript \\
        rails & 4 & 2 & 2 & \textasciitilde450k & Ruby \\
        phpipam & 4 & 2 & 2 & \textasciitilde225k & PHP \\
        assimp & 4 & 4 & 0 & \textasciitilde562k & C++ \\
        siyuan & 3 & 1 & 2 & \textasciitilde265k & Go, TypeScript \\
        cvat-ai & 3 & 3 & 0 & \textasciitilde320k & Py, TypeScript \\
        MobSF & 3 & 1 & 2 & \textasciitilde132k & JS, Py \\
        mlflow & 3 & 1 & 2 & \textasciitilde870k & Py \\
        \bottomrule
    \end{tabularx}
    \caption{Top 15 projects identified in the large-scale evaluation of \cvegenie, showing the number of reproduced CVEs, their classification into CWE Top 25 vs. other CWEs, lines of code (LoC), and primary languages.}
    \label{tab:projects}
\end{table}

\subsection{Large Scale Evaluation}
\label{subsec:large-scale}

\descr{Dataset $(D_L)$.} For our large-scale evaluation, we construct a dataset, $(D_L)$, consisting of all CVEs published between June 2024 and May 2025 that include publicly available source code. 
A CVE is included in $(D_L)$ if at least one of the following conditions is satisfied: (1) the CVE entry contains a GitHub URL (e.g., a patch commit) in its references that links directly to the relevant source code, or (2) the CVE specifies affected software versions, and the source code corresponding to those versions is publicly accessible.
Because these CVEs were disclosed after the knowledge cutoff of the LLMs (\texttt{o3} and \texttt{o4-mini}) used in \cvegenie, $(D_L)$ ensures zero prior exposure during model training.
Overall, $(D_L)$ comprises 841 CVEs spanning 186 CWEs, 29 programming languages, and 440 distinct open-source software (OSS) projects. 
The OSS projects cover a broad range of application domains, including web applications and backends (156), libraries and frameworks (147), CLI tools and binaries (28), cloud and DevOps systems (25), embedded and networking software (18), operating systems and runtimes (17), AI/ML platforms (16), desktop applications (13), blockchain and cryptocurrency systems (7), mobile applications and SDKs (7), and security tools (6).
In terms of vulnerability types, the majority of CVEs in $(D_L)$ fall under the following categories: injection and improper neutralization (29\%), authorization and authentication (11\%), resource management (8\%), information exposure (6\%), memory safety (3\%), request forgery (3\%), and others.
Our statistical analysis further shows that approximately 55\% of CVEs in $D_L$ fall under the top 25 most dangerous CWEs~\cite{mitre--top25_nodate}, and around 49\% lack a PoC in their advisory.

\descr{Methodology.} We use the same methodology as Section~\ref{subsec:sat-study}, and run \cvegenie{} on $D_L$ three times, iteratively.

\descr{Results.} \cvegenie{} reproduced \totalcves{} CVEs out of 841 (Figure~\ref{fig:large-scale}), spanning \totallangs{} programming languages, \totalcwes{} vulnerability types, and \totalprojects{} projects. To better understand outcomes, we perform statistical analysis on all reproduction runs as well as manually analyze 50 successful and 50 failed cases, and categorize results as follows:

\descrit{Domain-Dependent Effectiveness of \cvegenie{}.} We observe that \cvegenie{} shows the highest performance for library frameworks and AI/ML platforms categories, with the reproduction rates of 78\% and 58\%, excelling on CWEs such as CWE-79 (XSS), CWE-89 (SQLi), and CWE-284 (access control). Meanwhile, the success appears more moderate for web/backend applications (45\%), where logic-driven CWEs such as 200, 22, and 918 have lower success rates. \cvegenie{} struggles in binary-heavy domains such as desktop (18\%), blockchain (8\%), and mobile (29\%) applications, especially with memory-safety (e.g., CWE-122, 416) and crypto/validation (CWE-295, 327) flaws. 
A finer-grained breakdown by vulnerability type and project characteristics is provided in the Appendix, where Table~\ref{tab:cwe-success} reports reproduction success rates for the top CWEs. Meanwhile, Table~\ref{tab:projects} and Table~\ref{tab:repro-summary} further contextualize results by project scale, language, and domain. Finally, Tables~\ref{tab:cwe-aiml}–\ref{tab:cwe-security} detail domain-specific CWE distributions, illustrating how failure %

\descrit{Failures.} For $198$ cost/time overrun failures, \textasciitilde80\% were due to cost overruns, where the agent either got stuck during project builds (\textasciitilde41\%) or while generating exploits (\textasciitilde59\%). The majority of failures involved web application CVEs lacking either vulnerability details or setup documentation. For example, none of the CVEs in project \verb|discourse| built successfully since its \verb|README| only links to external documentation, which our agents cannot access, causing the exhaustion of their budgets in exploring the codebase.

\begin{researchbox}
\textbf{RQ4:} At scale, \cvegenie{} reproduced \totalcves{} / 841 CVEs, spanning \totalprojects{} projects, \totallangs{} programming languages, and \totalcwes{} CWEs, demonstrating broad applicability across ecosystems. Notably, 217 (51\%) reproduced CVEs fall within the MITRE top 25 most dangerous CWEs, and 231 (54\%) had no PoC available.
\end{researchbox}

\section{Limitations and Future Work}
\descrit{Vulnerabilities Involve UI Interactions.} Currently, \cvegenie{} only supports command-line interface (CLI) interactions. However, many web-based vulnerabilities require interaction with a graphical UI, making them difficult to trigger via CLI. Future work should explore integrating \cvegenie{} with UI environments, e.g., web browsers.

\descrit{Multimodal CVE Knowledge Curation.} Due to a lack of image and video processing support, \cvegenie{} struggles with CVEs where the PoC is provided as an image or video. Future work can address this limitation by integrating \cvegenie{} with multimodal LLMs. Additionally, leveraging external CVE resources through LLM-based deep search tools, beyond standard vulnerability databases, may also help improve reproduction success rates~\cite{mu_understanding_2018}.

\descrit{Over Critique.} As shown in Figures~\ref{subfig:setup-critic}, \ref{subfig:exploit-critic}, and \ref{subfig:verifier-critic}, all critics exhibit a lower TPR compared to their TNR. While this helps reduce false positives, it also results in missing some valid reproductions, particularly in cases where critics requested detailed setup, exploit, or verifier that the developer could not sufficiently provide.

\descrit{Cost.} Running \cvegenie{} costs approximately \avgcost{} per CVE, moreover, open-source models currently do not perform well. Therefore, to support further research, we release logs from all reproduction runs as a novel dataset to help the community fine-tune and improve local models' performance for CVE reproduction.

\section{Conclusion}
In this paper we presented \cvegenie{}, an automated framework designed to reproduce real-world vulnerabilities at scale. This has considerable benefit for downstream security research (see Appendix~\ref{sec:applications}), which suffers from a shortage of diverse high-quality data suitable for research and development of new tooling for automated vulnerability discovery and repair. \cvegenie{} successfully reproduced around \totalcvesper{} (\totalcves{}) of 841 CVEs, with on average \avgcost{} cost per CVE, across a vast variety of projects, CWEs, and programming languages. To the best of the authors' knowledge, no other framework is capable of such comprehensive results at this scale and efficiency.

\appendix

\section{Acknowledgments}
We gratefully acknowledge and thank Dave Aitel, Harold Nguyen, Xin Hu, and Ian Brelinsky (OpenAI), as well as Stijn Pletinckx and Lukas Dresel (University of California, Santa Barbara), for their valuable discussions and guidance that contributed to this work.

\section{Ethical Considerations}
\descrit{Safe Release of Reproduced CVEs’ Data.} For all of our studies, we only include the CVEs that have a public patch available on the National Vulnerability Database (NVD). This measure prevents the attackers from using our reproduced CVEs to exploit the stakeholders' systems, and also helps us to safely release the data of the CVEs reproduced in our experimental studies.

\descrit{Release of CVE-Genie’s Codebase.} The aim of releasing CVE-Genie’s codebase is to support defenders and solution developers by accelerating vulnerability triaging and tool development for vulnerability detection and patching. Although it is also possible for the attackers to leverage CVE-Genie to reproduce vulnerabilities in the real world, our aim is to level the playing field for defenders; specifically, to empower defenders to reproduce CVE reports to attain up-to-date and reproducible CVE environments as benchmark test cases. These benchmarks will then be used to rigorously test and enhance vulnerability detection and patching systems, thereby understanding their performance and ideally accelerating their response capabilities. Following the beneficence criterion of the Menlo Report~\cite{dittrich_menlo_2012}, we strongly believe that our system provides more benefits than potential harms to the security community and to society.

\section{Open Science}
\label{sec:open-science}

\descrit{\cvegenie{} Framework and Dataset.}
\url{https://github.com/BUseclab/cve-genie}.
This includes (1) source code of \cvegenie{} and detailed instructions on how to run it, (2) reproduction runs of all reproduced CVEs with their agent conversations/logs, intermediate artifacts, and final exploit and verifier scripts, and (3) a web application to visualize the CVE reproduction runs.

\descrit{Experiments Results.} 
\url{https://osf.io/dcej4/overview?view_only=81309bbc7e8c489abb9fb7e40bd1c0fb}.
This includes all reproduction runs with their agent conversations/logs, intermediate artifacts, and final scripts. (1) \verb|critics_evals.tar.gz| contains results for selecting optimal LLMs and prompts (Section 3.5). (2) \verb|baseline.tar.gz| contains results for the baseline assessment study (Section 4.1). (3) \verb|design_ablation.tar.gz| contains results for \cvegenie{}'s design ablations (Section 4.2). (4) \verb|robustness.tar.gz| contains results for \cvegenie{}'s robustness to loss of CVE context (Section 4.3). (5) \verb|large_scale.tar.gz| contains results for multiple runs of the large-scale study with 841 CVEs (Section 4.4).

\section{Generative AI Usage}
We used ChatGPT (OpenAI) to assist with LaTeX formatting, as well as to review and refine spelling, grammar, and sentence structure for improved clarity of the manuscript. Moreover,
all AI-assisted outputs were manually reviewed, edited as needed, and explicitly approved by the authors before being incorporated into the paper.

\section{\cvegenie{} Specifications}
\label{app:tools}
\cvegenie{} is built on LangChain and supports easy design and integration of tools for LLMs. For \cvegenie{} to effectively reproduce CVEs, it require access to a project's source code. However, due to LLMs' limited context windows, large project files cannot be processed all at once. To address this, we developed primitives enabling directory browsing using ``\emph{project directory tree}'' and command execution:

\begin{compactenum}
    \item \textbf{get\_file:} The LLM agent specifies the file path and the number of lines it wants to read from a given offset. To prevent context overflow and improve efficiency, it reads a maximum of 300 lines at a time. If more content is needed, the agent can scroll up or down to access it.

    \item \textbf{write\_to\_file:} The LLM provides the path and content for a file, which we then write to the specified file. This facilitates modifications and updates within the project.

    \item \textbf{execute\_ls\_command:} The LLM specifies a directory, and we execute the \texttt{ls} command within it, returning the output. This command is distinct from \texttt{execute\_linux\_command} due to its major role in \emph{project directory} exploration and its frequent use.

    \item \textbf{execute\_linux\_command:} The LLM issues Linux commands to be executed from the project's root directory. Foreground commands have a 300-second timeout, with standard output and error logged to separate files. After completion, we return up to the last 100 lines of the log and the log file path which LLM can read using \verb|get_file| for further analysis. Background execution is supported for non-blocking tasks (e.g., starting a web server). For these, we show output after 5 seconds, provide the log file path, and indicate whether the process is still running.

    \item \textbf{set\_environment\_variable:} This tool allows for setting environment variables for subsequent \texttt{execute\_linux\_command} calls. We observed that LLMs frequently issue calls to export environment variables, which are not retained in successive runs as each command is executed in a separate shell. To address this, we enable the LLM to set these variables, which are then passed to commands. There is also an option to clear all set variables.
\end{compactenum}

\descr{Restrictions.} Each tool-calling agent is limited to 60 tool calls. This limit was found to be optimal in preliminary experiments for maximizing CVE reproduction across all LLMs.

\descr{Output Formatting.} For all agents requiring output in a specific format, we add a \verb|gpt-4o-mini|-based format corrector to the output parser. If parsing fails, the raw output is passed to \verb|gpt-4o-mini| for formatting. If formatting fails after three attempts, an error is returned.

\begin{table}[h]
    \centering
    \resizebox{0.5\textwidth}{!}{
    \begin{tabular}{r|l|l}
    \toprule
        \textbf{Category} & \textbf{Application} & \textbf{Benefits brought by \cvegenie{}}  \\ 
        \hline
        \multirow{3}{*}{\begin{tabular}[r]{@{}r@{}}Vul. \\ detection\end{tabular}} 
        &  ML-based detection & Training and testing data \\ \cline{2-3}
        &  Dynamic fuzzing & Benchmarks and initial fuzzing seeds \\ \cline{2-3}
        &  Static analysis & Benchmarks and static rules \\ \hline
        \multirow{2}{*}{\begin{tabular}[r]{@{}r@{}}Vul. \\ triage\end{tabular}}
        &  Root cause analysis & Enrich inputs and enable more tools \\ \cline{2-3}
        &  Exploit generation  & Ingredients for exploit chaining  \\ \hline
        \multirow{2}{*}{\begin{tabular}[r]{@{}r@{}}Vul. \\ Patching\end{tabular}}
         &  Patch generation & Benchmarks and testing environments \\ \cline{2-3}
         &  Patch verification & Inputs and environments\\ \hline
        \multirow{2}{*}{\begin{tabular}[r]{@{}r@{}} Other \\ applications\end{tabular}} 
        & LLM secure code generation & Benchmarks and testing environment  \\ \cline{2-3}
        &  Penetration testing & Construct pen. test tasks and environment  \\ \cline{2-3}
        &  Attack detection & Benchmarks and detection rules  \\ \bottomrule
    \end{tabular}}
    \caption{Tasks that benefit from \cvegenie{}.}
    \label{tab:app}
\end{table}

\section{Applications of \cvegenie}
\label{sec:applications}

As shown in Table~\ref{tab:app}, reproducible CVEs can be used for a plethora of security analysis tasks, including  vulnerability detection, triage, exploitation, and patching.

\descr{Vulnerability Detection.}
Most existing vulnerability detection datasets are built using patch commits from CVEs (see Table \ref{tab:datasets}). They typically label the pre-patch version of a function as vulnerable and the post-patch version as benign~\cite{nikitopoulos_crossvul_2021, fan_bigvul_2020, bhandari_cvefixes_2021}. However, extracting isolated functions often strips away critical context~\cite{risse_topscore_2024}, and functions labeled for a single vulnerability may contain others, resulting in noisy labels. Using reproducible CVEs with working PoCs can address these issues. By executing PoCs during function code extraction, we can ensure that vulnerable execution traces are included, preserving necessary context and confirming the presence of vulnerabilities. Additionally, this approach supports both static and dynamic analysis, unlike most existing datasets (e.g., SVEN~\cite{he_sven_2023}, BigVul~\cite{fan_bigvul_2020}, PrimeVul~\cite{ding_vulnerability_2024}), which are limited to static methods.

\descr{Vulnerability Triage and Patching.}
Reproducible CVEs are essential for effective vulnerability triage. They help eliminate false positives and provide an executable environment with known vulnerable inputs, which aids in accurately analyzing and reproducing bugs. This capability enables advanced techniques like reverse execution and backward taint analysis for root cause identification. \cvegenie{} supports this process by offering exploits for individual vulnerabilities, which can be combined into complex exploit chains for studying multi-stage attacks. It also serves as a benchmark for evaluating patching methods, providing rich contextual data such as crash stack traces and outputs. Additionally, its PoCs assist in verifying and refining patches. 

\descr{LLM Insecure Code Generation.}
Reproducible CVEs are valuable for evaluating the security of code generated by LLMs. Recent studies have shown that LLMs can produce insecure code, particularly in security-critical contexts~\cite{yang_seccodeplt_2024, wan_cyberseceval_2024}. Various benchmarks evaluate secure code generation capability of LLM, such as, CyberSecEval~\cite{wan_cyberseceval_2024} prompts an LLM to generate code snippets and uses rule-based detectors to identify insecure code, however, rule-based detection often leads to false positives. To improve this, SecCodePLT~\cite{yang_seccodeplt_2024} leverages CWE data and manually crafts dynamic test cases to evaluate secure code generation. Reproducible CVEs offer a scalable alternative because these CVEs are confirmed to be vulnerable, they can be used to assess LLMs' ability to generate secure code. By prompting the LLM to implement the functionalities in the vulnerable functions, and using the associated PoCs to test the output, we can determine whether the generated code is vulnerable.

\descr{Penetration Testing and Attack Detection.}
\cvegenie{} offers a valuable resource for both penetration testing and cybersecurity training. It provides ready-to-use exploitation scenarios along with verified solutions, making it ideal for educating future penetration testers and developing automated testing tools.
For defenders, these exploits can be analyzed to identify attack patterns, which can enhance detection mechanisms such as intrusion detection systems (IDS) or malware classifiers.

\section{Single Monolithic Agent Study Design}
\label{sec:single-agent}
To evaluate a standalone LLM without \cvegenie{}’s modular, agentic design and structured guidance, we collapsed all agents into a single monolithic agent with access to the full toolset (Appendix A), its entire conversation history, and an increased limit of 100 tool calls per attempt. The agent was provided with all CVE-related information directly in the user prompt, and the system prompt explicitly instructed it to follow a three-stage workflow (project setup, exploit generation, and verification), mirroring \cvegenie{}’s high-level procedure. Unlike \cvegenie{}, no automated critic or feedback loops were used; instead, we manually inspected execution traces and the generated exploit and verifier artifacts to determine whether reproduction was successful.

\section{Case Study for CVE-2024-5129}
\label{sec:case-study-cve-2024-5129}

\descr{Overview.}
Versions of the \texttt{lunary} project prior to v1.2.8 expose a \verb|DELETE /v1/datasets/:id| endpoint that lacks authorization checks and performs an unscoped deletion by dataset UUID. In the vulnerable release v1.2.7, the handler executes a SQL \verb|DELETE| using only \verb|id|, with no authentication and no restriction to the caller’s project, allowing unauthenticated deletion of arbitrary datasets by guessing or obtaining UUIDs. 

\begin{table}[]
    \centering
    \footnotesize
    \begin{tabularx}{\linewidth}{
        l
        >{\centering\arraybackslash}X
        >{\centering\arraybackslash}X
        >{\centering\arraybackslash}X
        >{\centering\arraybackslash}X
        >{\centering\arraybackslash}X
    }
        \toprule
        \textbf{CWE-IDs} & \textbf{\# CVEs} & \multicolumn{2}{c}{\textbf{PoC Available}} & \multicolumn{2}{c}{\textbf{No PoC}} \\
        \cmidrule(lr){3-4}\cmidrule(lr){5-6}
        & & \textbf{Reprod} & \textbf{Failed} & \textbf{Reprod} & \textbf{Failed} \\
        \midrule
        CWE-502 & 6 & 3 & 1 & 1 & 1 \\
        CWE-29  & 3 & 2 & 1 & 0 & 0 \\
        CWE-20  & 3 & 1 & 2 & 0 & 0 \\
        CWE-77  & 3 & 0 & 2 & 1 & 0 \\
        CWE-94  & 2 & 0 & 1 & 0 & 1 \\
        \bottomrule
    \end{tabularx}
    \caption{Top CWEs reproduced in AI/ML Platform projects.}
    \label{tab:cwe-aiml}
\end{table}

\begin{table}[]
    \centering
    \footnotesize
    \begin{tabularx}{\linewidth}{
        l
        >{\centering\arraybackslash}X
        >{\centering\arraybackslash}X
        >{\centering\arraybackslash}X
        >{\centering\arraybackslash}X
        >{\centering\arraybackslash}X
    }
        \toprule
        \textbf{CWE-IDs} & \textbf{\# CVEs} & \multicolumn{2}{c}{\textbf{PoC Available}} & \multicolumn{2}{c}{\textbf{No PoC}} \\
        \cmidrule(lr){3-4}\cmidrule(lr){5-6}
        & & \textbf{Reprod} & \textbf{Failed} & \textbf{Reprod} & \textbf{Failed} \\
        \midrule

        CWE-79  & 99 & 34 & 37 & 10 & 18 \\
        CWE-284 & 29 & 12 &  8 &  2 &  7 \\
        CWE-22  & 21 &  7 &  6 &  3 &  5 \\
        CWE-200 & 19 &  5 &  3 &  2 &  9 \\
        CWE-918 & 16 &  4 &  5 &  1 &  6 \\
        CWE-89  & 15 &  5 &  6 &  1 &  3 \\
        CWE-400 & 15 &  6 &  3 &  1 &  5 \\
        CWE-20  & 12 &  4 &  3 &  1 &  4 \\
        CWE-285 &  9 &  5 &  2 &  2 &  0 \\

        \bottomrule
    \end{tabularx}
    \caption{Top CWEs reproduced in Web App projects.}
    \label{tab:cwe-web}
\end{table}

\begin{table}[]
    \centering
    \footnotesize
    \begin{tabularx}{\linewidth}{
        l
        >{\centering\arraybackslash}X
        >{\centering\arraybackslash}X
        >{\centering\arraybackslash}X
        >{\centering\arraybackslash}X
        >{\centering\arraybackslash}X
    }
        \toprule
        \textbf{CWE-IDs} & \textbf{\# CVEs} & \multicolumn{2}{c}{\textbf{PoC Available}} & \multicolumn{2}{c}{\textbf{No PoC}} \\
        \cmidrule(lr){3-4}\cmidrule(lr){5-6}
        & & \textbf{Reprod} & \textbf{Failed} & \textbf{Reprod} & \textbf{Failed} \\
        \midrule
        CWE-79   & 33 &  9 & 6 & 12 & 6 \\
        CWE-1333 & 15 &  8 & 1 &  4 & 2 \\
        CWE-400  & 13 &  5 & 1 &  6 & 1 \\
        CWE-200  & 13 &  9 & 0 &  1 & 3 \\
        CWE-770  & 13 &  5 & 0 &  5 & 3 \\
        CWE-22   &  8 &  3 & 1 &  2 & 2 \\
        \bottomrule
    \end{tabularx}
    \caption{Top CWEs reproduced in Lib/Framework projects.}
    \label{tab:cwe-lib}
\end{table}

\begin{table}[]
    \centering
    \footnotesize
    \begin{tabularx}{\linewidth}{
        l
        >{\centering\arraybackslash}X
        >{\centering\arraybackslash}X
        >{\centering\arraybackslash}X
        >{\centering\arraybackslash}X
        >{\centering\arraybackslash}X
    }
        \toprule
        \textbf{CWE-IDs} & \textbf{\# CVEs} & \multicolumn{2}{c}{\textbf{PoC Available}} & \multicolumn{2}{c}{\textbf{No PoC}} \\
        \cmidrule(lr){3-4}\cmidrule(lr){5-6}
        & & \textbf{Reprod} & \textbf{Failed} & \textbf{Reprod} & \textbf{Failed} \\
        \midrule
        CWE-285 & 4 & 0 & 1 & 0 & 3 \\
        CWE-209 & 3 & 0 & 2 & 0 & 1 \\
        CWE-200 & 3 & 0 & 0 & 1 & 2 \\
        CWE-269 & 2 & 1 & 1 & 0 & 0 \\
        CWE-287 & 2 & 0 & 0 & 0 & 2 \\
        \bottomrule
    \end{tabularx}
    \caption{Top CWEs reproduced in Cloud/DevOps projects.}
    \label{tab:cwe-cloud}
\end{table}

\begin{table}[]
    \centering
    \footnotesize
    \begin{tabularx}{\linewidth}{
        l
        >{\centering\arraybackslash}X
        >{\centering\arraybackslash}X
        >{\centering\arraybackslash}X
        >{\centering\arraybackslash}X
        >{\centering\arraybackslash}X
    }
        \toprule
        \textbf{CWE-IDs} & \textbf{\# CVEs} & \multicolumn{2}{c}{\textbf{PoC Available}} & \multicolumn{2}{c}{\textbf{No PoC}} \\
        \cmidrule(lr){3-4}\cmidrule(lr){5-6}
        & & \textbf{Reprod} & \textbf{Failed} & \textbf{Reprod} & \textbf{Failed} \\
        \midrule
        No-ID    & 7 & 0 & 0 & 5 & 2 \\
        CWE-122  & 3 & 0 & 0 & 3 & 0 \\
        CWE-400  & 2 & 0 & 0 & 1 & 1 \\
        CWE-416  & 2 & 0 & 0 & 0 & 2 \\
        CWE-863  & 2 & 0 & 0 & 0 & 2 \\
        \bottomrule
    \end{tabularx}
    \caption{Top CWEs reproduced in OS/Runtime projects.}
    \label{tab:cwe-os}
\end{table}

\begin{table}[]
    \centering
    \footnotesize
    \begin{tabularx}{\linewidth}{
        l
        >{\centering\arraybackslash}X
        >{\centering\arraybackslash}X
        >{\centering\arraybackslash}X
        >{\centering\arraybackslash}X
        >{\centering\arraybackslash}X
    }
        \toprule
        \textbf{CWE-IDs} & \textbf{\# CVEs} & \multicolumn{2}{c}{\textbf{PoC Available}} & \multicolumn{2}{c}{\textbf{No PoC}} \\
        \cmidrule(lr){3-4}\cmidrule(lr){5-6}
        & & \textbf{Reprod} & \textbf{Failed} & \textbf{Reprod} & \textbf{Failed} \\
        \midrule
        CWE-79  & 7 & 0 & 6 & 0 & 1 \\
        CWE-122 & 4 & 0 & 0 & 1 & 3 \\
        CWE-416 & 4 & 0 & 0 & 2 & 2 \\
        CWE-22  & 2 & 0 & 2 & 0 & 0 \\
        CWE-20  & 2 & 0 & 1 & 0 & 1 \\
        \bottomrule
    \end{tabularx}
    \caption{Top CWEs reproduced in Desktop App projects.}
    \label{tab:cwe-desktop}
\end{table}

\begin{table}[]
    \centering
    \footnotesize
    \begin{tabularx}{\linewidth}{
        l
        >{\centering\arraybackslash}X
        >{\centering\arraybackslash}X
        >{\centering\arraybackslash}X
        >{\centering\arraybackslash}X
        >{\centering\arraybackslash}X
    }
        \toprule
        \textbf{CWE-IDs} & \textbf{\# CVEs} & \multicolumn{2}{c}{\textbf{PoC Available}} & \multicolumn{2}{c}{\textbf{No PoC}} \\
        \cmidrule(lr){3-4}\cmidrule(lr){5-6}
        & & \textbf{Reprod} & \textbf{Failed} & \textbf{Reprod} & \textbf{Failed} \\
        \midrule
        No-ID   & 4 & 0 & 0 & 1 & 3 \\
        CWE-345 & 3 & 2 & 0 & 1 & 0 \\
        CWE-116 & 2 & 0 & 0 & 1 & 1 \\
        CWE-147 & 2 & 0 & 0 & 1 & 1 \\
        CWE-150 & 2 & 0 & 0 & 1 & 1 \\
        \bottomrule
    \end{tabularx}
    \caption{Top CWEs reproduced in CLI Tool/Utility projects.}
    \label{tab:cwe-cli}
\end{table}

\begin{table}[]
    \centering
    \footnotesize
    \begin{tabularx}{\linewidth}{
        l
        >{\centering\arraybackslash}X
        >{\centering\arraybackslash}X
        >{\centering\arraybackslash}X
        >{\centering\arraybackslash}X
        >{\centering\arraybackslash}X
    }
        \toprule
        \textbf{CWE-IDs} & \textbf{\# CVEs} & \multicolumn{2}{c}{\textbf{PoC Available}} & \multicolumn{2}{c}{\textbf{No PoC}} \\
        \cmidrule(lr){3-4}\cmidrule(lr){5-6}
        & & \textbf{Reprod} & \textbf{Failed} & \textbf{Reprod} & \textbf{Failed} \\
        \midrule
        CWE-670  & 3 & 1 & 1 & 0 & 1 \\
        CWE-200  & 2 & 0 & 2 & 0 & 0 \\
        CWE-400  & 1 & 0 & 1 & 0 & 0 \\
        \bottomrule
    \end{tabularx}
    \caption{Top CWEs reproduced in Blockch/Crypto projects.}
    \label{tab:cwe-crypto}
\end{table}

\begin{table}[]
    \centering
    \footnotesize
    \begin{tabularx}{\linewidth}{
        l
        >{\centering\arraybackslash}X
        >{\centering\arraybackslash}X
        >{\centering\arraybackslash}X
        >{\centering\arraybackslash}X
        >{\centering\arraybackslash}X
    }
        \toprule
        \textbf{CWE-IDs} & \textbf{\# CVEs} & \multicolumn{2}{c}{\textbf{PoC Available}} & \multicolumn{2}{c}{\textbf{No PoC}} \\
        \cmidrule(lr){3-4}\cmidrule(lr){5-6}
        & & \textbf{Reprod} & \textbf{Failed} & \textbf{Reprod} & \textbf{Failed} \\
        \midrule
        CWE-532 & 2 & 0 & 0 & 1 & 1 \\
        CWE-261 & 1 & 0 & 0 & 1 & 0 \\
        CWE-327 & 1 & 0 & 0 & 1 & 0 \\
        \bottomrule
    \end{tabularx}
    \caption{Top CWEs reproduced in MobileApp/SDK projects.}
    \label{tab:cwe-mobile}
\end{table}

\begin{table}[]
    \centering
    \footnotesize
    \begin{tabularx}{\linewidth}{
        l
        >{\centering\arraybackslash}X
        >{\centering\arraybackslash}X
        >{\centering\arraybackslash}X
        >{\centering\arraybackslash}X
        >{\centering\arraybackslash}X
    }
        \toprule
        \textbf{CWE-IDs} & \textbf{\# CVEs} & \multicolumn{2}{c}{\textbf{PoC Available}} & \multicolumn{2}{c}{\textbf{No PoC}} \\
        \cmidrule(lr){3-4}\cmidrule(lr){5-6}
        & & \textbf{Reprod} & \textbf{Failed} & \textbf{Reprod} & \textbf{Failed} \\
        \midrule
        CWE-200 & 4 & 1 & 0 & 1 & 2 \\
        CWE-416 & 3 & 0 & 0 & 1 & 2 \\
        CWE-670 & 2 & 1 & 0 & 0 & 1 \\
        CWE-639 & 2 & 1 & 0 & 0 & 1 \\
        \bottomrule
    \end{tabularx}
    \caption{Top CWEs reproduced in Emb/Network projects.}
    \label{tab:cwe-embedded}
\end{table}

\begin{table}[]
    \centering
    \footnotesize
    \begin{tabularx}{\linewidth}{
        l
        >{\centering\arraybackslash}X
        >{\centering\arraybackslash}X
        >{\centering\arraybackslash}X
        >{\centering\arraybackslash}X
        >{\centering\arraybackslash}X
    }
        \toprule
        \textbf{CWE-IDs} & \textbf{\# CVEs} & \multicolumn{2}{c}{\textbf{PoC Available}} & \multicolumn{2}{c}{\textbf{No PoC}} \\
        \cmidrule(lr){3-4}\cmidrule(lr){5-6}
        & & \textbf{Reprod} & \textbf{Failed} & \textbf{Reprod} & \textbf{Failed} \\
        \midrule
        CWE-918  & 3 & 0 & 2 & 0 & 1 \\
        CWE-863  & 2 & 0 & 0 & 2 & 0 \\
        CWE-23   & 1 & 1 & 0 & 0 & 0 \\
        \bottomrule
    \end{tabularx}
    \caption{Top CWEs reproduced in Security/Server projects.}
    \label{tab:cwe-security}
\end{table}

\descr{\cveprocessor.}
The workflow starts with \dataprocessor{} resolving CVE-2024-5129’s affected versions and automatically downloading the latest vulnerable release (v1.2.7) from the upstream repository. \dataprocessor{} also extracts the CVE data, CWE information, patch commit, and bug bounty report for the CVE. The \knowledge then constructs a structured CVE knowledge base summarizing the vulnerability as missing authorization (CWE-862) and localizing the root cause to \verb|v1/datasets/index.ts|, where the vulnerable delete handler deletes by dataset ID alone. The knowledge base also records the upstream fix (adding authorization middleware and scoping deletion by \verb|projectId|), which becomes the ground truth for later exploit and verifier synthesis.

\descr{\project.}
Given the vulnerable source tree and knowledge base, the \project{} reconstructs a runnable environment. The \predev{} agent inspects repository structure and reads key configuration files (e.g., root \verb|package.json|, backend \verb|package.json|, backend \verb|.env.example|, migration runner \verb|src/migrate.ts|) to infer the required service stack: a Node.js/Koa backend backed by PostgreSQL and initialized via SQL migrations under \verb|packages/db|.

The \setupdev{} agent then provisions a real PostgreSQL 14 service inside the VM, configures credentials, and generates a valid \verb|.env| file pointing \verb|DATABASE_URL| to the local database instance. All Node.js dependencies are installed via \verb|npm install|, and database migrations are executed using the project’s native migration runner (\verb|npm run migrate:db|), resulting in a fully initialized schema. Finally, the backend is launched using \verb|npm run dev| and confirmed operational by observing successful database connectivity and a listening API server on port 3333.
The \setupcritic{} agent analyzes the execution logs of \setupdev{} and approves the build.

After environment reconstruction, the \setupcritic{} evaluates the entire setup. The critic confirms that (i) the vulnerable code path is present and unchanged, (ii) a real PostgreSQL-backed Lunary backend is running, (iii) no dummy services or mocked endpoints were introduced, and (iv) the vulnerable \verb|DELETE /v1/datasets/:id| endpoint is reachable and responds with HTTP 200 to unauthenticated requests. Based on these checks, the critic accepts the project build as a genuine and exploitable reproduction target.

\descr{\exploit.}
The \exploitdev{} agent begins by performing targeted code inspection to ground the exploit in the real implementation. It reads the vulnerable dataset router source file and confirms that the \verb|datasets.delete("/:id")| handler executes an unscoped SQL deletion without any authentication or authorization middleware. It also inspects backend configuration files to verify the correct runtime entry points and dependencies.
To validate exploitability in practice, the exploiter launches the backend and issues unauthenticated \verb|DELETE| requests against \verb|/v1/datasets/<uuid>| using both ad-hoc Python test scripts and raw HTTP calls. A randomly generated UUID is sufficient to elicit an unconditional HTTP 200 response, confirming that no permission checks or project scoping are enforced.
Based on this confirmation, the exploiter generates a standalone PoC \verb|exploit.py| script. The script accepts a dataset UUID (and an optional base URL), sends an unauthenticated \verb|DELETE| request to the vulnerable endpoint, and prints the HTTP status code and response body. The exploit is intentionally minimal, containing no setup logic or auxiliary assumptions beyond network access to the API.

The generated exploit artifact is then evaluated by the \exploitcritic agent. The critic verifies that the exploit (i) targets the correct vulnerable endpoint, (ii) directly triggers the missing-authorization flaw described in the CVE knowledge base, (iii) does not rely on environmental shortcuts or fabricated behavior, and (iv) has been empirically validated against the running vulnerable service. After confirming these properties, the critic accepts the exploit as a correct and faithful PoC.

\descr{\verifier.}
To automatically confirm successful exploitation of CVE-2024-5129, the \verifierdev{} agent synthesizes a CTF-style verifier that executes the exploit and validates its effect against the reconstructed \verb|Lunary| backend. The verifier follows a fixed the following three-phase structure. 

\begin{compactenum}
    \item In the \emph{pre-setup} phase, it inserts a minimal but valid set of rows, i.e., organization, account, project, and dataset, directly into the PostgreSQL database used by the running backend. A fresh dataset UUID is generated and recorded, ensuring that a concrete deletion target exists prior to exploitation.

    \item In the \emph{exploit} phase, the verifier invokes the previously generated \texttt{exploit.py} script verbatim via \texttt{subprocess.run}, passing the dataset UUID and the base URL \url{http://localhost:3333}. The verifier itself performs no HTTP requests, ensuring that the exploit remains the sole trigger of the vulnerability.

    \item In the \emph{post-exploit} phase, the verifier re-queries the database to determine whether the dataset row with UUID still exists. If the row has been deleted, the verifier prints the flag \texttt{3xploit66full}; otherwise, it reports failure. Database access is implemented robustly, using \texttt{psycopg2} when available and falling back to the \texttt{psql} client if necessary.
\end{compactenum}

During verification, earlier verifier iterations that relied on mocked or stubbed HTTP servers were rejected by the \verifiercritic{} agent. The final accepted verifier operates exclusively on the real Node.js / Koa backend and PostgreSQL instance and performs a direct state-based validation. The \flagcheck{} successfully executes the verifier, observes flag emission, and the \verifiercritic{} confirms the verifier as a correct and faithful reproduction of the CVE.

\descr{\emph{Storing Artifacts and Metadata}.} After the successful verification, \cvegenie{} stores the VM snapshot with running \verb|Lunary| server. Moreover, it stores exploit and verifier scripts, agent logs, and metadata of the reproduction run. Overall, this reproduction run of CVE-2024-5129 cost \$1.68 and 20 minutes. See artifacts here \footnote{\url{https://github.com/BUseclab/cve-genie/tree/main/results/reproduced_cves/CVE-2024-5129}} in open-source dataset.

\bibliographystyle{plain}
\bibliography{ref}

\end{document}